\documentstyle[sprocl,epsf]{article}

\def\be{\begin{equation}}
\def\ee{\end{equation}}
\def\bea{\begin{eqnarray}}
\def\eea{\end{eqnarray}}

\newcommand{\bef}{\begin{figure}}
\newcommand{\eef}{\end{figure}}
\newcommand{\hmp}{ h^{-1}Mpc}
\newcommand{\etal}{{\it et al.}}
\def\spose#1{\hbox to 0pt{#1\hss}}
\def\ltapprox{\mathrel{\spose{\lower 3pt\hbox{$\mathchar"218$}}
 \raise 2.0pt\hbox{$\mathchar"13C$}}}
\def\gtapprox{\mathrel{\spose{\lower 3pt\hbox{$\mathchar"218$}}
 \raise 2.0pt\hbox{$\mathchar"13E$}}}
\def\inapprox{\mathrel{\spose{\lower 3pt\hbox{$\mathchar"218$}}
 \raise 2.0pt\hbox{$\mathchar"232$}}}

\begin{document}

\title{Correlation properties of the large scale matter  
distribution and galaxy 
 number counts}

\author{L. Pietronero, F. Sylos Labini and  M. Montuori}

\address{Dipartimento di Fisica, Universit\`a di Roma
``La Sapienza''\\
P.le A. Moro 2, I-00185 Roma, Italy \\
and INFM sezione di Roma 1}

\maketitle
\abstracts{ We introduce the basic techniques
used for the analysis
 of three dimensional and two dimensional galaxy samples. 
We report the correlation analysis of various redshift surveys 
which shows that the available data are consistent with each other 
and manifest fractal correlations (with dimension $D \simeq 2$) 
up to the present observational limits  
without any tendency towards homogenization.
This result points to a new interpretation of 
the galaxy number counts.
We show that an analysis of the small scale fluctuations allows us 
to reconcile the correlation analysis and the number counts in a 
new perspective which has a number of important implications.
}
  
\section{Introduction}

In this lecture  we briefly introduce the basics concepts 
of fractal geometry and the 
methods of correlation analysis, that are usually 
used in Statistical Mechanics. First we introduce
the methods to compute the full correlation function in three
dimensional samples characterized by a wide solid angle.
then we consider the determination of the 
space density in very deep and narrow surveys.
Such an analysis allows us to study and characterize
the effect of the small scale fluctuations in the 
determination of non-average quantities, as 
the radial density. Having clarified these
effects we move to the interpretation of
the galaxy number counts, i.e. the number of galaxies
with a certain apparent magnitude versus the magnitude itself.
These counts must be examined with great care, because, also in this 
case, it is not possible to perform an average over different observers.
We show that an analysis of the small scale fluctuations allows us 
to reconcile the correlation analysis and the number counts in a 
new perspective which has a number of important implications.

Finally  we summarize our main conclusions.
For a more detailed discussion we refer the 
reader to Coleman \& Pietronero(1992) and Baryshev \etal 
(1994) for a basic introduction to this 
approach, and to Pietronero \etal (1997) and 
Sylos Labini \etal (1997) for a review on the more recent results
(see Davis (1997) for a different point of view on this subject).


\section{Space distribution}

We first introduce some basic definitions.
If $L$ is the absolute or intrinsic
luminosity of a galaxy at distance $r$,
this   appears with an apparent flux
\be
\label{e41}
f = \frac{L}{4 \pi r^2}  \; .
\ee
For historical reasons the apparent magnitude $m$
of an object with incoming flux $f$
is 
\be
\label{e42}
m = -2.5 \log_{10}f + constant,
\ee
while the absolute magnitude $M$
 is instead related to its intrinsic luminosity $L$ by
\be
\label{e43}
M=-2.5 \log_{10}L + constant'.
\ee
From Eq.\ref{e41} it follows that
the difference between the apparent and the absolute magnitudes
 of an
object at distance $r$ is (at relatively small distances,
neglecting relativistic effects)
\be
\label{e44}
m-M = 5 \log_{10}r + 25
\ee
where $r$ is expressed in Megaparsec ($1 Mpc= 3 \cdot 10^{24}
 \; cm$).

A catalog is usually obtained by measuring the redshifts 
of the all galaxies with apparent magnitude brighter
than a certain apparent magnitude limit $m_{lim}$,
 in a certain region of the sky defined by a solid angle $\Omega$.
An important selection effect exists, in that at every distance in the 
apparent magnitude limited survey, there is a definite limit in intrinsic 
luminosity
which is the absolute magnitude of the fainter galaxy
which can be seen at that distance. Hence at large distances, intrinsically
faint objects are not observed whereas at smaller distances they are observed.
In order to analyze the statistical properties of galaxy distribution,
a catalog which does not suffer for 
this selection effect
must be used. In general, it exists a very well known procedure to
obtain a sample that is not biased by this luminosity selection effect:
this is the so-called {\it  "volume limited"} (VL) sample.
A  VL  sample contains every galaxy in the volume
which is more luminous than a certain limit, so that in such a
 sample
there is no incompleteness for an observational
luminosity selection effect \cite{dp83,cp92}.
Such a sample is defined by a certain maximum distance $R_{VL}$
and the absolute magnitude limit $M_{VL}$ 
given by
\be
\label{e45}
M_{VL}=m_{lim}-5\log_{10}R_{VL} -25 -A(z)
\ee
where $A(z)$ takes into account various corrections (K-corrections, 
absorption, relativistic effects, etc.), and 
$m_{lim}$ is the survey apparent magnitude limit.
 Different VL samples, 
extracted from one catalog, have different $R_{VL}$, and 
deeper is a VL sample, larger its $R_{VL}$.

The measured velocities of the galaxies have been
expressed in
the preferred frame of the Cosmic Microwave Background Radiation
(CMBR), i.e. the heliocentric velocities of
   the galaxies have been corrected
for the solar motion with respect to the CMBR, according
 with the formula
\be
\label{e46}
\vec{v} =\vec{v}_{m}+316 cos \theta \; \; km s^{-1}
\ee
where $\vec{v}$ is the corrected velocity,
$\vec{v}_{m}$ is the observed velocity and $\theta$
is the angle between the observed
velocity and the direction of the CMBR dipole anisotropy
($\alpha=169.5^{\circ}$ and $\delta=-7.5^{\circ}$).
From these corrected velocities, we have calculated
 the comoving distances
$r(z)$, with for example $q_0=0.5$, by using the Mattig's relation
\cite{bslmp94}
\be
\label{e47}
r(z)=6000\left(1-\frac{1}{\sqrt{(1+z)}}\right) \hmp \; .
\ee
In general, we have checked that the results of our analysis
depend very weakly on the particular value of $q_0$ adopted, 
except very deep surveys, and
we have also used the simple linear relation
\be
\label{e48}
r = cz/H_{0} \; .
\ee
In the nearby catalogs 
there is no any sensible change by using Eq.\ref{e48} 
instead of Eq.\ref{e47}.
Hereafter for the Hubble constant 
 we use the value 
$H_{0}=100\cdot  h \cdot km sec^{-1} Mpc^{-1}$. 

All the analyses presented here have been 
performed in redshift space and we have not
adopted {\it any correction} to take into account the 
eventual effect of peculiar velocities 
(local distortion to the Hubble flow).
However we point out that peculiar velocities have an 
amplitude up to $\sim 500 \div 1000 km sec^{-1}$ 
and then their effect can be important only up
to $ 5 \div 10 \hmp$, and not more (see the lecture of 
A. Szaley in these Proceedings)

We briefly mention the characteristics 
of the {\it galaxy luminosity distribution}
that is useful in what follows 
analyses. The basic assumption 
we use to  compute all the following
 relations
is that:
\be
\label{e49}
\nu(L,\vec{r}) = \phi(L) \rho(\vec{r}) \; , 
\ee
i.e. that the number of galaxies for unit luminosity
and volume $\nu(L,\vec{r})$ can be expressed as
the product of the space density
$ \rho(\vec{r})$  and  the luminosity function $ \phi(L)$
($L$ is the intrinsic luminosity).
This is a crude approximation
in view of the multifractal properties of the distribution
(correlation between position and luminosity) \cite{slp96}.
However, for the purpose of the present discussion,
the approximation of Eq.\ref{e49} is rather good and the
explicit consideration of the multifractal properties
 have a minor effect on the properties we  
discuss \cite{slp96}.
                    
To each VL sample (limited by the 
absolute magnitude $M_{VL}$)
we can associate the luminosity factor
\be
\label{e410}
\Phi(M_{VL}) = \int_{-\infty}^{M_{VL}} \phi(M) dM  
\ee
that gives the fraction of galaxies 
present in the sample. 
Hereafter we adopt the following normalization
for the luminosity function
\be
\label{e411}
\Phi(\infty) = \int_{-\infty}^{M_{min}} \phi(M) dM  = 1 
\ee
where $M_{min} \approx -10 \div -12$ is the fainter galaxy present in 
the available samples. The luminosity factor of Eq.\ref{e410} is useful to
normalize the space density in different VL samples which 
have different $M_{VL}$ (Eq.\ref{e45}). The 
luminosity function measured in real catalogs has 
the so-called Schecther like shape
\cite{sch76}
\be
\label{e412}
\phi(M)dM = A \cdot 10^{-0.4(\delta+1)M} e^{-10^{0.4(M^*-M)}}dM
\ee
where  $\delta \approx -1.1$ and $M^* \approx -19.5$ \cite{dac94},
\cite{vet94}, and the constant $A$ is given by the normalization
 condition of Eq.\ref{e411}. 

Eq.\ref{e49} and eq.\ref{e410} are  useful for the normalization of the 
average (conditional) density 
in different VL samples. In fact, in view of eq.\ref{e49},
the difference of the average density in different VL samples,
 is only due to the luminosity factor given by eq.\ref{e410}.


\subsection{Full correlation  analysis}
    
In this section we mention the essential properties     
of fractal structures because they will be necessary for    
the correct interpretation of the statistical analysis.     
However in no way these properties are assumed    
 or used in the analysis itself.    

A fractal consists of a system in which more     
and more structures appear at smaller and     
smaller scales and the structures at small     
scales are similar to the ones at large scales.    
The first quantitative description of these forms is     
the metric dimension.    
One way to determine it, is the     
computation of mass-length    
relation. Starting from an     
 point occupied by an object     
of the distribution, we count how     
many objects $N(r)$    
("mass") are present, in average, within a volume     
of linear size  $r$ ("length")     
\cite{man83}:    
\be    
\label{l1}    
<N(r)> = B\cdot r^{D}    
\ee    
 $D$ is the fractal dimension     
and characterises in a quantitative way    
 how the system fills the space.    
The prefactor $\:B$     
depends to the lower cut-offs of the distribution; these    
are related to the smallest scale above     
which the system is self-similar     
and below which the self similarity     
is no more satisfied.    
In general we can write:    
\be    
\label{l6}    
B = \frac {N_{*}} {{r_{*}}^{D}}    
\ee    
where $r_{*}$ is this smallest scale     
and $N_{*}$ is the number of object    
up to $r_{*}$.    
For a deterministic fractal this relation is exact, while     
for a stochastic one  it is satisfied in an average sense.    
Eq.(\ref{l1}) corresponds to a average behaviour     
of $N(r)$, that is a very fluctuating     
function; a fractal is, in fact, characterised by  large    
 fluctuations and clustering at all scales.    
We stress that eq.(\ref{l1}) is completely general, i.e. it holds also     
for an homogeneous distribution, for which $D=3$.     
From eq.(\ref{l1}), we can compute the     
average density $\:<n>$ for a sample of    
 radius $\:R_{s}$ which contains a portion     
of the structure with dimension $D$.     
Assuming for simplicity a spherical volume     
($\:V(R_{s}) = (4/3)\pi R_{s}^{3}$), we have    
\be    
\label{l2}    
<n> =\frac{N(R_{s})}{V(R_{s})} = \frac{3}{4\pi } B R_{s}^{-(3-D)}    
\ee    
If the distribution is homogeneous ($D = 3$)   the    
average density is constant and independent from the sample     
volume; in the case of  a fractal, the average density     
 depends explicitly on the sample     
size $\:R_{s}$ and it is not a meaningful     
quantity. In particular, for a fractal     
the average density is a decreasing function of the sample size and     
$<n> \rightarrow 0$ for $ R_{s} \rightarrow \infty$.    
    
It is important to note that eq.(\ref{l1})     
holds from every point of the     
system, when considered as the origin.    
This feature is related to the non-analyticity of the     
distribution.    
In a fractal every observer is     
equivalent to any other one, i.e. it holds the property     
of local isotropy around any observer     
\cite {sl94}.

The first quantity able to analyze the spatial properties     
of point distributions is the average density.    
Coleman \& Pietronero (1992) 
proposed     
the conditional density defined as:    
\be    
\label{g2}    
\Gamma(r) = \frac{<n(\vec{r}+\vec{r}_{i})    
n(\vec{r}_{i})>_{i}}{<n>}     
\ee    
where the index $i$ means that the average     
is performed over the points $r_{i}$     
of the distribution.   
In other words, we consider 
spherical volumes of radius $r$ around   
each points  of the sample and we measure   
the average density of points inside them.    
Such spherical volumes have to be   
fully contained in the sample boundaries. In Eq.\ref{g2}  
$\:<n>$ is the average density of the sample;     
this normalisation does not introduce any bias even if the average    
density is sample-depth dependent,     
as in the case of fractal distributions,     
as one can see from Eq. \ref{g3}.    
$\Gamma(r)$ (Eq. \ref{g2})    
can be computed by the following expression    
\begin{eqnarray}    
\label{g3}    
\Gamma(r)& =\frac{1}{M(r)} \sum_{i=1}^{M(r)}     
\frac{1}{4 \pi r^2 \Delta r}     
\int_{r}^{r+\Delta r} n(\vec{r}_i+\vec{r'})d\vec{r'}=\nonumber\\    
& &= \frac{DB}{4 \pi}  r^{3-D}    
\end{eqnarray}
where $M(r)$ is the number of points which contribute 
which are at distance  $\gtapprox r$ from any of the 
boundaries of the sample.    
$\Gamma(r)$ is a smooth function away from the     
lower and upper cutoffs of the distribution ($r_{*}$ and     
the dimension of the sample).    
From Eq.(\ref{g3}), we can see that     
$\Gamma(r)$ is independent from the     
sample size, depending only by     
the intrinsic quantities of the distribution     
($B$ and $D$). 
If the sample is homogeneous,$D=3$,     
$\:\Gamma(r) $     
is constant.    
If the sample is fractal, then 
$D < 3$, $\:\gamma > 0$ and     
$\Gamma(r)$ is a {\it power law}.    
For a more complete discussion we refer the reader to     
\cite{cp92}, \cite{slmp97}.    
If the distribution is fractal up     
to a certain distance $\lambda_0$,    
and then it becomes homogeneous,    
we have that:    
$$    
\Gamma(r) = \frac{BD}{4 \pi} r^{D-3} \;  r < \lambda_0    
$$    
\be    
\label{e327b}     
\Gamma(r)= \frac{BD}{4 \pi} \lambda_0^{D-3} \; r \geq \lambda_0    
\ee    
    
It is also very useful to use the {\it conditional average density}    
defined as:     
\be    
\label{g4}    
\Gamma^*(r) = \frac{3}{4 \pi r^3} \int_{0}^{r} 4 \pi r'^2 \Gamma(r') dr'    
\ee    
This function produce an artificial smoothing of     
$\Gamma(r)$ function,     
but it correctly     
reproduces global properties \cite{cp92}.    
     
Given a certain spherical sample with solid angle $\Omega$ and depth $R_s$,
it is important to define which is 
 the maximum distance up to which it 
is possible to compute the correlation function ($\Gamma(r)$ or $\xi(r)$).
As discussed in Sylos Labini \etal (1997), 
we   limit our analysis to an
effective  depth
$R_{eff}$ that is of the order of the radius of the maximum
sphere fully contained in the sample volume (Fig.\ref{fig10}).
\bef  
\epsfxsize 4cm
\centerline{\epsfbox{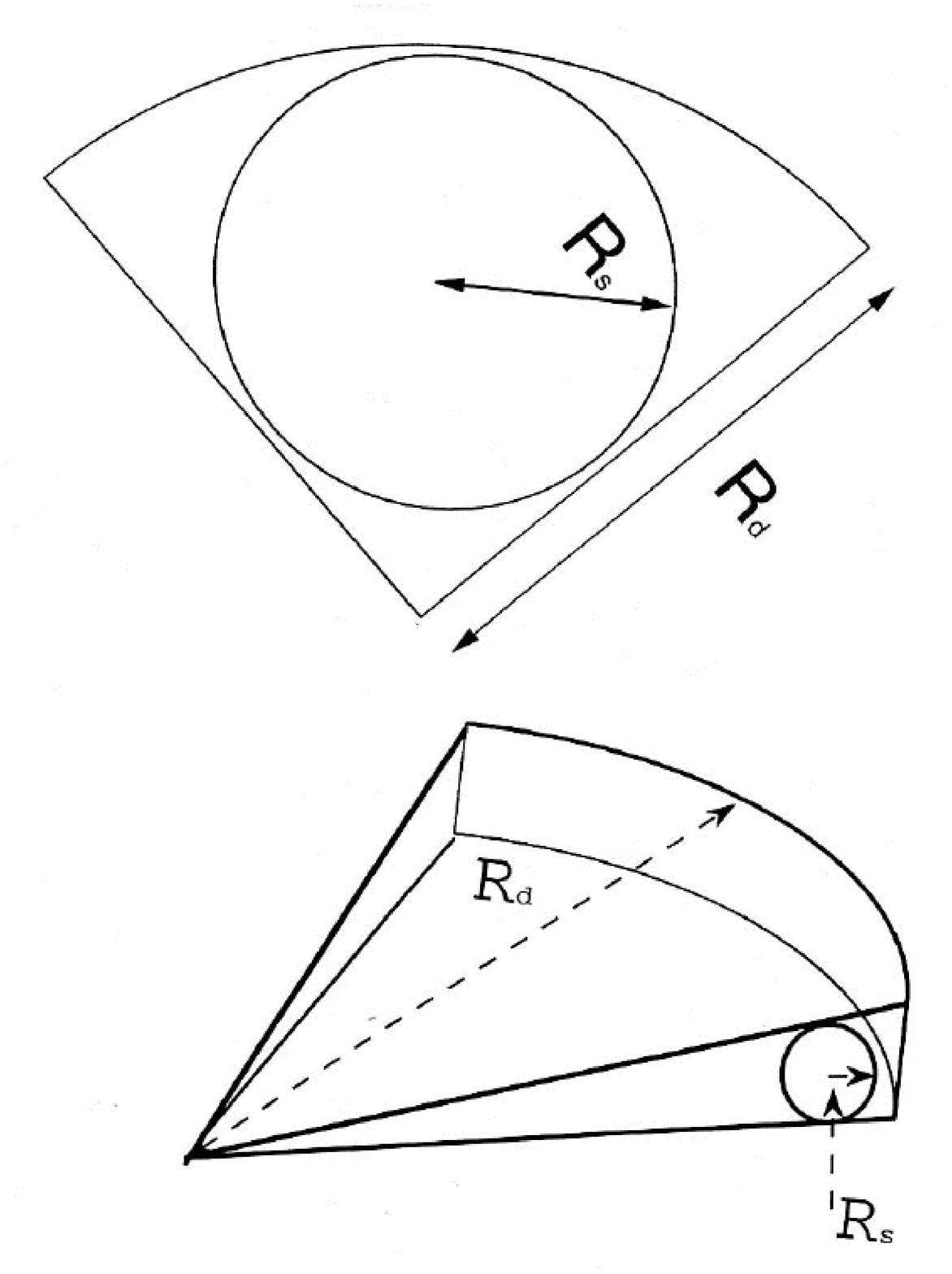}}
\caption{{\it Upper part:} A typical cone diagram for a wide angle
galaxy catalog 
(e.g. CfA, SSRS, Perseus-Pisces).
  The depth is $R_d$. The {\it effective depth}
 is $R_{eff}$ 
and it corresponds to the radius of the maximum
 sphere fully contained in the sample volume 
 ($R_{eff} \ltapprox R_s$).
{\it Bottom part:} A typical cone diagram for 
a narrow angle galaxy catalog (e.g. LCRS, ESP).   
In this case $R_{eff} << R_d$. 
\label{fig10}}
\eef
For example for a catalog with the limits in right ascension
($\alpha_1 \leq \alpha \leq \alpha_2$) and declination
($\delta_1 \leq \delta \leq \delta_2$) we have that 
\be
\label{ws1}
R_{eff} = \frac{R_s sin (\delta \theta /2)}{1+sin(\delta \theta/2)}
\ee
where $\delta \theta = min (\alpha_2 - \alpha_1, \delta_2-\delta_1)$.
In such a way  we do not consider in the statistics
the points for which a sphere of radius {\it r} is not
fully included within the sample boundaries.
Hence we do not make use of any weighting scheme
with the advantage of not making 
any assumption in the treatment of
the boundaries conditions.
For this reason we have a smaller number of points
and we stop our analysis  at a  smaller depth than that
of other authors.

 The reason why
$\Gamma(r)$ (or $\xi(r)$ see in what follows) cannot
be computed for $r > R_{eff}$
is essentially the following.
 When one evaluates the correlation
function
(or power spectrum) beyond $R_{eff}$,
then one  makes explicit assumptions on what
lies beyond the sample's boundary.  In fact, even in absence of
corrections for selection effects, one
is forced to consider incomplete shells
calculating $\Gamma(r)$ for $r>R_{eff}$,
thereby
implicitly assuming that what one  does not find  in the part of the
shell not included in the sample is equal to what is inside (or other
similar weighting schemes).
In other words, the standard calculation
introduces a spurious homogenization which we are trying to remove.

If one could reproduce via an analysis that uses weighting schemes, the 
correct properties of the distribution under analysis, it would be
not necessary to produce wide angle survey, and from a single pencil beam
deep survey it would be possible to study the entire matter
distribution up to very deep scales. It is evident that
this could not be the case.
By the way, we have done  a test on the homogenization effects
of weighting schemes on artificial distributions as well as on
real catalogs, finding that the flattening of the 
conditional density
 is indeed introduced owing to  the weighting,
and does not correspond to any real feature in the galaxy distribution
\cite{slmp97}.

\subsection{Standard analysis}  
  
At this point it is instructive to   
consider the behaviour of the standard   
correlation function $\xi(r)$.  
Coleman \& Pietronero (1992) clarify some crucial   
points of the  
such an analysis, and in particular they discuss the meaning  
of the so-called {\it "correlation length"}  
  $\:r_{0}$  
found with the standard  
approach \cite{dp83} \cite{pee93}
and defined by the relation:  
\be  
\label{x1}  
\xi(r_{0})= 1  
\ee  
where  
\be  
\label{x2}  
\xi(r) = \frac{<n(\vec{r_{i}})n(\vec{r_{i}}  
+ \vec{r})>_{i}}{<n>^{2}}-1  
\ee  
is the two point correlation function used in the   
standard analysis.  
If the average density is not a well defined intrinsic   
property of the system, the analysis with  
$\xi(r)$ gives spurious results.  
In particular, if the system has   
fractal correlations, the average density is simply related   
to the sample size as shown by Eq.(\ref{l2}).  
In other words, it is meaningless to   
define the correlation length   
of the distribution by comparing the average   
correlation $<n(\vec{r_{i}})n(\vec{r_{i}}  
+ \vec{r})>_{i}$ to the average density   
of the sample $<n>^2$, if   
the latter depends on the sample volume.  
The expression of the   
$\:\xi(r)$ for a 
fractal distribution is \cite{slmp97} 
\be  
\label{x3}  
\xi(r) = ((3-\gamma)/3)(r/R_{s})^{-\gamma} -1  
\ee  
where $\:R_{s}$ (the effective sample   
radius) is the radius of   
the spherical volume where one computes the  
average density from Eq. (\ref{l2}).  
From Eq. (\ref{x3}) it follows that:  
       i.) the so-called correlation  
length $\:r_{0}$ (defined as $\:\xi(r_{0}) = 1$)  
is a linear function of the sample size $\:R_{s}$  
\be  
\label{x4}  
r_{0} = ((3-\gamma)/6)^{\frac{1}{\gamma}}R_{s}  
\ee  
and hence it is a quantity without any correlation 
meaning, but it is  
simply related to the sample size.  
  
ii.) the amplitude of the $\xi(r)$ is:  
\be   
\label  {x5}  
A(R_{s}) = ((3-\gamma)/3)R_{s}^{\gamma}   
\ee

iii.) $\:\xi(r)$ is a power law only for   
\be  
\label{x6}  
((3-\gamma)/3)(r/R_{s})^{-\gamma}  >> 1  
\ee  
hence for $\: r \ltapprox r_{0}$: for larger distances  
there is a clear deviation from a  
power law behavior due to the definition of $\:\xi(r)$.  
This deviation, however, is just due to the size of  
 the observational sample and does not correspond to any real change  
of the correlation properties. It is clear that if one estimates the  
 exponent of $\xi(r)$ at distances $r \gtapprox r_0$, one  
 systematically obtains a higher value of the correlation exponent  
 due to the break of $\xi(r)$ in the log-log plot.   
This is actually the case for the analyses performed so far:  
 in fact, usually,   
$\xi(r)$ is fitted with a power law in the   
range $ 0.5 r_{0} \ltapprox    
r \ltapprox 2 r_{0}$, where we get an higher value of   
 the correlation  exponent. In particular,    
the usual estimation of this exponent  
by the  $\xi(r)$ function leads to $\gamma \approx 1.7$, different   
from $\gamma \approx 1$ (corresponding  
to $D \approx 2$), that we found by means of    
the $\Gamma(r)$ analysis \cite{slmp97}.


\subsection{Normalization of the average density in different surveys}

We have already 
defined the luminosity factor which  is associated
to each VL sample.   As long as the
 space and the luminosity density can be considered independent,
 the normalization of $\Gamma(r)$  in different VL
 samples can be simply done by dividing their amplitudes for the
 corresponding luminosity factors. 
Of course such a normalization is parametric, because it depends on
 the two parameters of the luminosity function $\delta$ and $M^*$
For a reasonable choice of these two
 parameters we find that the amplitude of the conditional and radial
 density matches quite well in different VL samples. In particular:
\begin{itemize}

\item (1) For CfA1, PP, ESP, LEDA and APM the parameters of the
 luminosity function are $\delta= -1.1$ and $M^* = -19.5$. All these
 surveys have been selected in the B-band, and hence the luminosity
 function is the same. 

\item (2) For SSRS1 the selection criteria have been chosen in the
 apparent diameter $d$, and hence we use the linear-diameter    
 function,
 rather than the luminosity function for the normalization of the
 amplitude in different VL samples. We use the diameter function:
\be
\label{diamfunc}
N(D)dD=N_0\cdot exp(-\delta_d \cdot D)dD
\ee
where $D$ is the absolute diameter (in Kpc), 
$\delta_d=0.109$ and $N_0$ is a constant \cite{pat95}.
The shape of this linear-diameter function is similar to 
the Schecther one for magnitudes.

\item (3) Galaxies in LCRS have been measured in the $r$ band, so
 that we have used the luminosity function given by Schetman \etal (1996),
  i.e. a Schecther like function with parameters
$\delta= -0.9$ and $M^* = -20.03$.

\item (4) The two IRAS surveys have been measured in the near
 infrared. In this case we have normalized the amplitude of the
 density using the $IR$ luminosity function  \cite{sau90}.
\end{itemize}

We find that in all the cases (1)-(4) the density amplitude  
 in the different VL samples match quite well 
(see Figs.\ref{fig73}-\ref{fig74}). 
\bef 
\epsfxsize 5cm 
\centerline{\epsfbox{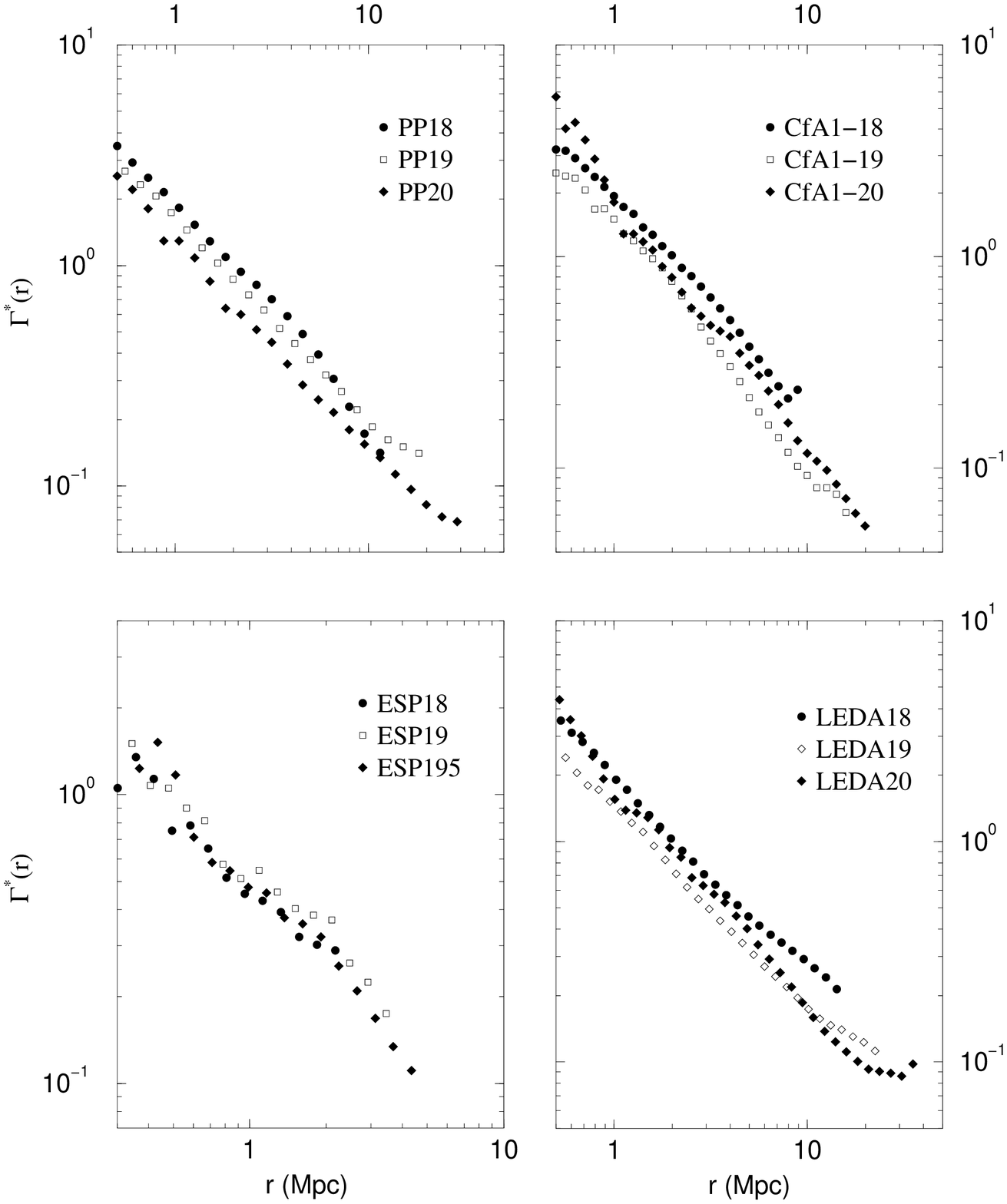}}  
\caption{\label{fig73} 
The spatial conditional average density 
$\Gamma^*(r)$ computed in some 
VL samples of Perseus-Pisces, CfA1, ESP and LEDA
and normalized to the corresponding
luminosity factor. } 
\eef 
\bef 
 \epsfxsize 5cm 
\centerline{\epsfbox{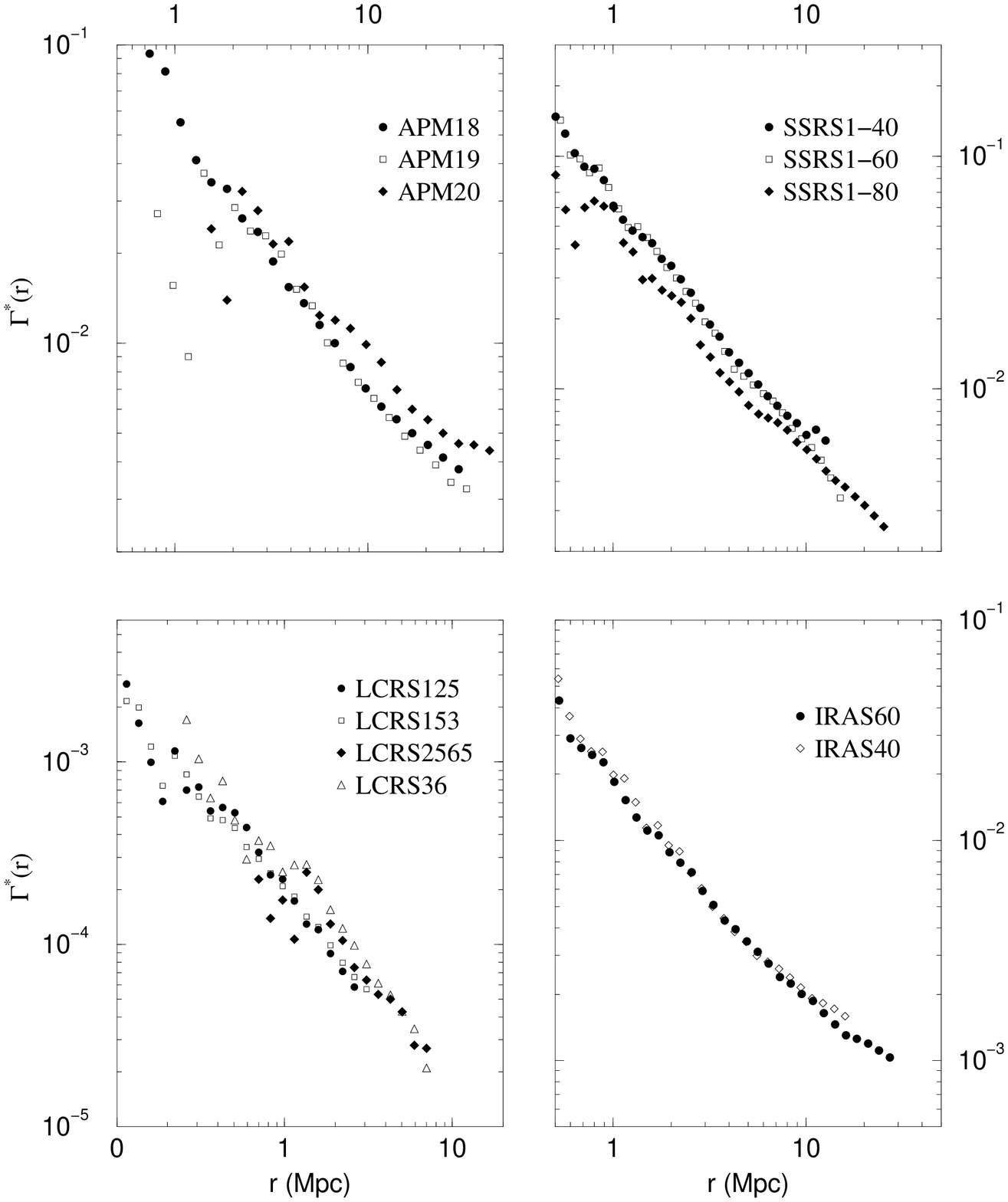}}  
\caption{\label{fig74} 
The spatial conditional average density $\Gamma^*(r)$ 
computed in some 
VL samples of APM, LCRS, SSRS1, IRAS $1.2 Jy$ 
and normalized to the 
luminosity factor. } 
\eef

The main data of our correlation analysis 
are collected in Fig.\ref{fig82} (left part)
 in which we report the 
{\it conditional density as a function of scale}
 for the various catalogues. 
The relative position of the various lines is not arbitrary but it is fixed 
by the luminosity function, a part for the cases of 
IRAS and SSRS1 for which this is 
not possible. The properties derived from different 
catalogues are compatible with each other and show a {\it power law 
decay} for the conditional density from $1 \hmp$ to $150 \hmp$
 without any tendency towards homogenization (flattening). This 
implies necessarily that the value of $r_0$ (derived from the $\xi(r)$ 
approach) will scale with the sample size $R_s$ as shown also from the 
specific data about $r_0$ of the various catalogues. The 
behaviour 
observed  corresponds to a fractal structure with dimension $D \simeq 
2$. (The data beyond $\sim 200 \hmp$ 
have been obtained by measuring the radial density 
and are therefore weaker from a statistical point of view
- see below)
The smaller value of CfA1 was due to its limited size. An 
homogeneous distribution would correspond to a flattening of the 
conditional density which is never observed
\bef
 \epsfxsize 9cm 
\centerline{\epsfbox{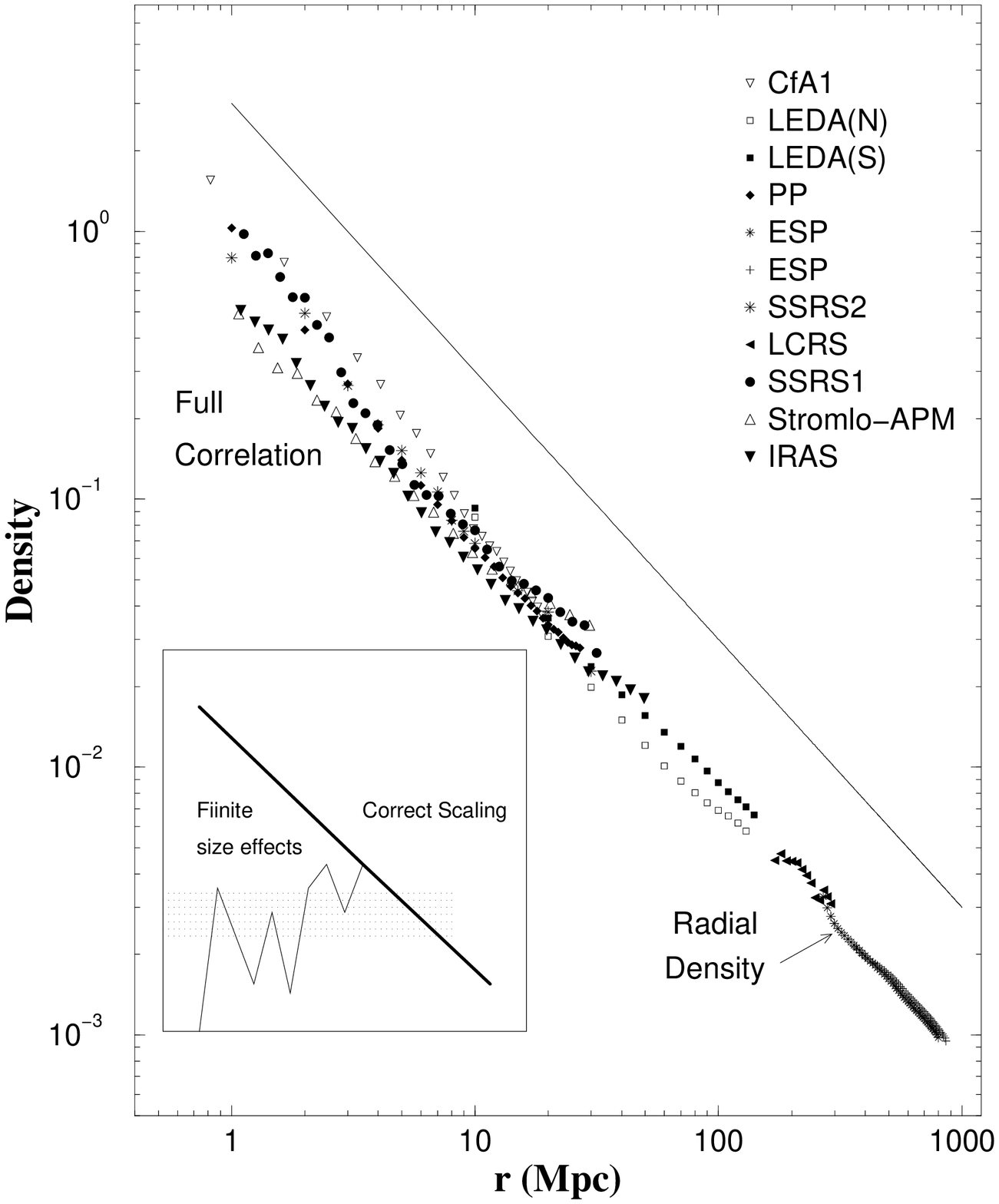}}  
\caption{\label{fig82} Full correlation analysis for the various
 available redshift surveys in the range of distance $0.5 \div 1000
 \hmp$. A reference line with slope $-1$ is also shown, that 
corresponds to fractal dimension $D = 2$. 
 } 
\eef 
It is
 remarkable to stress that the amplitudes and the slopes of the
 different surveys match quite well. From this figure we conclude
 that galaxy correlations show very well defined fractal properties
 in the entire range $0.5 \div 1000 \hmp$ with dimension $D = 2 \pm
 0.2$. Moreover all the surveys are in agreement with each other.

 It is interesting to compare the analysis of Fig.\ref{fig82} with 
the usual one, made with the function $\xi(r)$, for the same 
galaxy catalogs. This is reported in Fig.\ref{fig83}
\bef
\epsfxsize 8cm 
\centerline{\epsfbox{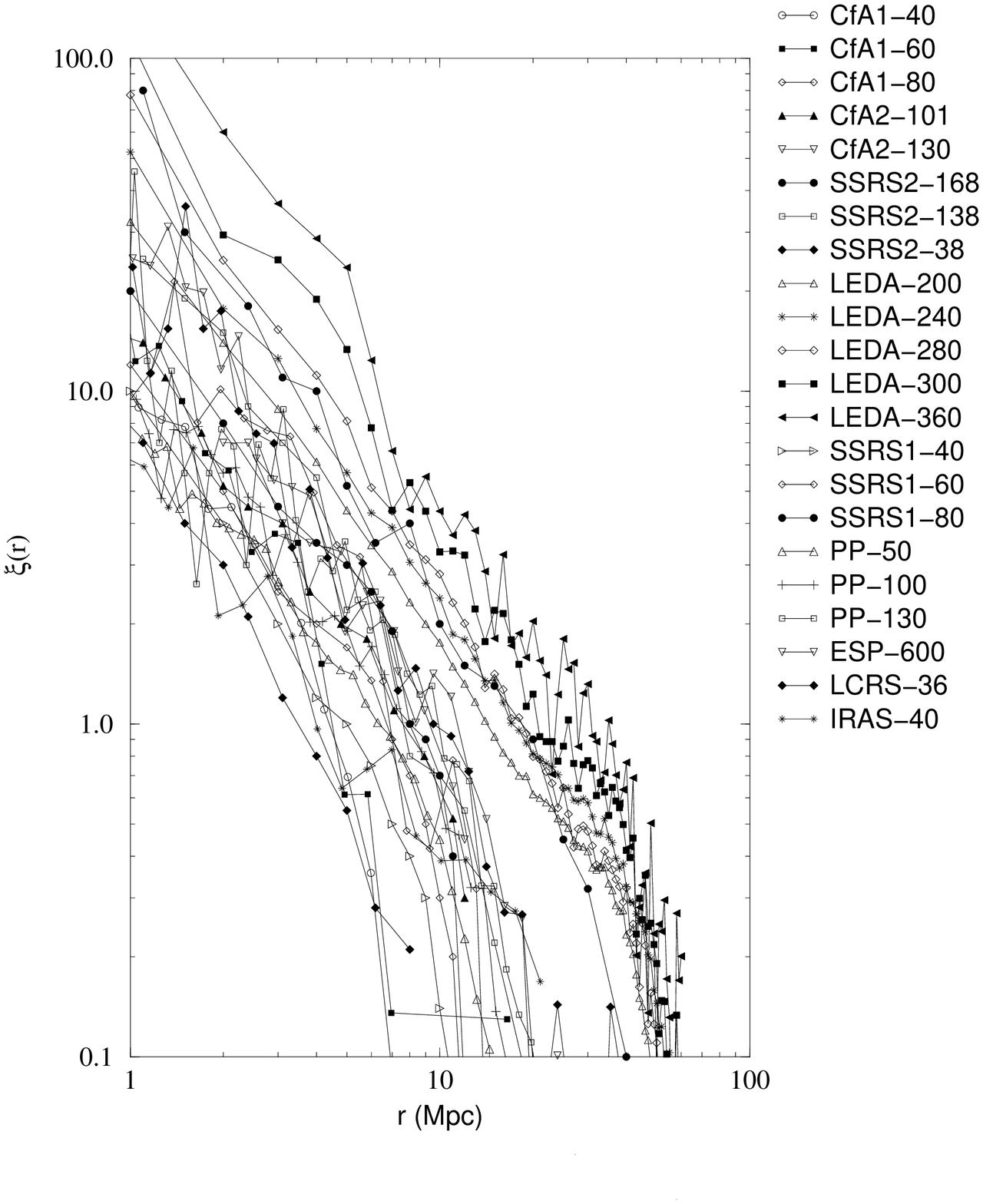}}  
\caption{\label{fig83}
Traditional analyses based on the function $\xi(r)$
of the same galaxy catalogs of the previous figure.
 The usual 
analysis is based on the a priori untested assumptions of 
analyticity and homogeneity. These properties
are not present in the real galaxy distribution and 
the results appear therefore rather confusing. 
This lead to the impression that galaxy catalogs are not good
enough and to a variety of theoretical problems like the 
galaxy-cluster mismatch, luminosity segregation, linear and 
non-linear evolution, etc.. This situation changes completely and 
becomes quite clear if one adopts the more 
general conceptual framework that is at the basis 
the previous figure}
\eef 
and, from this point of view, the various data 
the various data appear to be in strong disagreement with 
each other. This is due to the fact that the usual analysis
looks at the data from the prospective of analyticity and large
scale homogeneity (within each sample). These properties have never
been tested and they are not present in the real galaxy
distribution so the result is rather confusing (Fig.\ref{fig83}).
Once the same data are analyzed with a broader perspective the
situation becomes clear (Fig.\ref{fig82}) and the data of 
different catalogs result in agreement with each other. It is 
important to remark that analyses like those of Fig.\ref{fig83}
have had a profound influence in the field in various ways: 
first the different catalogues appear in conflict with each other.
This has generated the concept of {\it not fair samples} and a 
strong mutual criticism about the validity of the data 
between different authors. In the other cases the
discrepancy observed in Fig.\ref{fig83} have been 
considered real physical problems for which various technical
approaches have been proposed. These problems are, for example, 
the galaxy-cluster mismatch, luminosity segregation, 
the richness-clustering relation and 
 the linear non-linear evolution of the perturbations 
corresponding to the {\it "small"} or  {\it "large"}
amplitudes of fluctuations. We can now see that all this problematic
situation is not real and it arises only from a 
statistical analysis based on inappropriate and too restrictive
 assumptions that do not find any correspondence in the 
physical reality. It is also important to note that,
even if the galaxy distribution would eventually became
homogeneous at larger scales, the use of the above statistical
concepts  is anyhow inappropriate for the range of scales 
in which the system shows fractal correlations as those 
shown in Fig.\ref{fig82}.


\section{Radial density}

In the previous sections we have discussed the methods
that allow one to measure the conditional (average) density
in real galaxy surveys. This statistical quantity is an average
 one, since it is determined by performing an average over all the
 points of the sample. We have discussed   the robustness
 and the limits of such a measurement. We have pointed out 
 that the estimate of the conditional density can be done up to
 a distance $R_{eff}$ which is of the order of the radius of the
 maximum sphere fully contained in the sample volume. This is
 because the conditional density must be computed {\it only 
in spherical shells.}
This condition puts a great limitation to the volume studied,
 especially in the case of 
deep and narrow surveys, for which the maximum depth $R_{s}$ 
can be 
one order of magnitude, or more, than the effective depth $R_{eff}$.

We discuss here the measurement of the {\it radial density} 
in VL samples \cite{slmp97}. 
The determination of such a quantity
allow us to extend the analysis of the space density well
 beyond the depth   $R_{eff}$.
The price to pay is that such a measurement is strongly affected by
 finite size spurious fluctuations, {\it because it is not an average
 quantity. } These finite size effects require a great caution 
 \cite{slgmp96}: the behaviour of statistical quantities (like the 
 radial density and the counts of galaxies as a function of the apparent
 magnitude) that are not averaged out, present new and subtle problems.

\subsection{Finite size effects and the behavior of the radial
 density} 
\label{radialfinite}

  In this section we discuss the general problem of the minimal
 sample size which is able to provide us with a statistically 
 meaningful information. For example, the mass-length relation for
 a   fractal, which defines the fractal dimension, is 
 \be 
\label{efs1} 
D=\lim_{r \rightarrow \infty} 
\frac{\log(N(<r))}{\log(r)} 
\ee  
However this relation is  {\em properly defined only in the
 asymptotic limit}, because only in  this limit  the fluctuations
 of  fractal structures are  self-averaging.  A fractal
 distribution is characterized by large fluctuations at all scales 
 and these fluctuations determine  the statistical properties of
 the structure. If the structure has a lower cut-off, as it is the
 case for any  real fractal, one needs a {\em "very large sample"}
 in order to recover the statistical properties of the distribution
 itself. Indeed, in any real physical problem  we would like to
 recover the asymptotic properties from the knowledge of a  {\em
 finite portion} of a fractal and the problem is that   a single
 finite realization of a random fractal is affected by finite size
 fluctuations due to the  lower cut-off.  

In a   homogeneous distribution we can define, in  average,
 a characteristics volume associated to each particle. This is the
 Voronoi volume \cite{vo08} $v_v$ whose radius $\ell_v$ is of the
 order of the mean particle  separation. It is clear that the
 statistical properties of the system can be defined only in
 volumes much larger than $v_v$. Up to this volume in fact we
 observe essentially nothing. Then one begins to include a few
 (strongly fluctuating) points, and finally, the correct scaling
 behavior is recovered  (Fig.\ref{fig61}). 
\bef 
\epsfxsize 12cm 
\centerline{\epsfbox{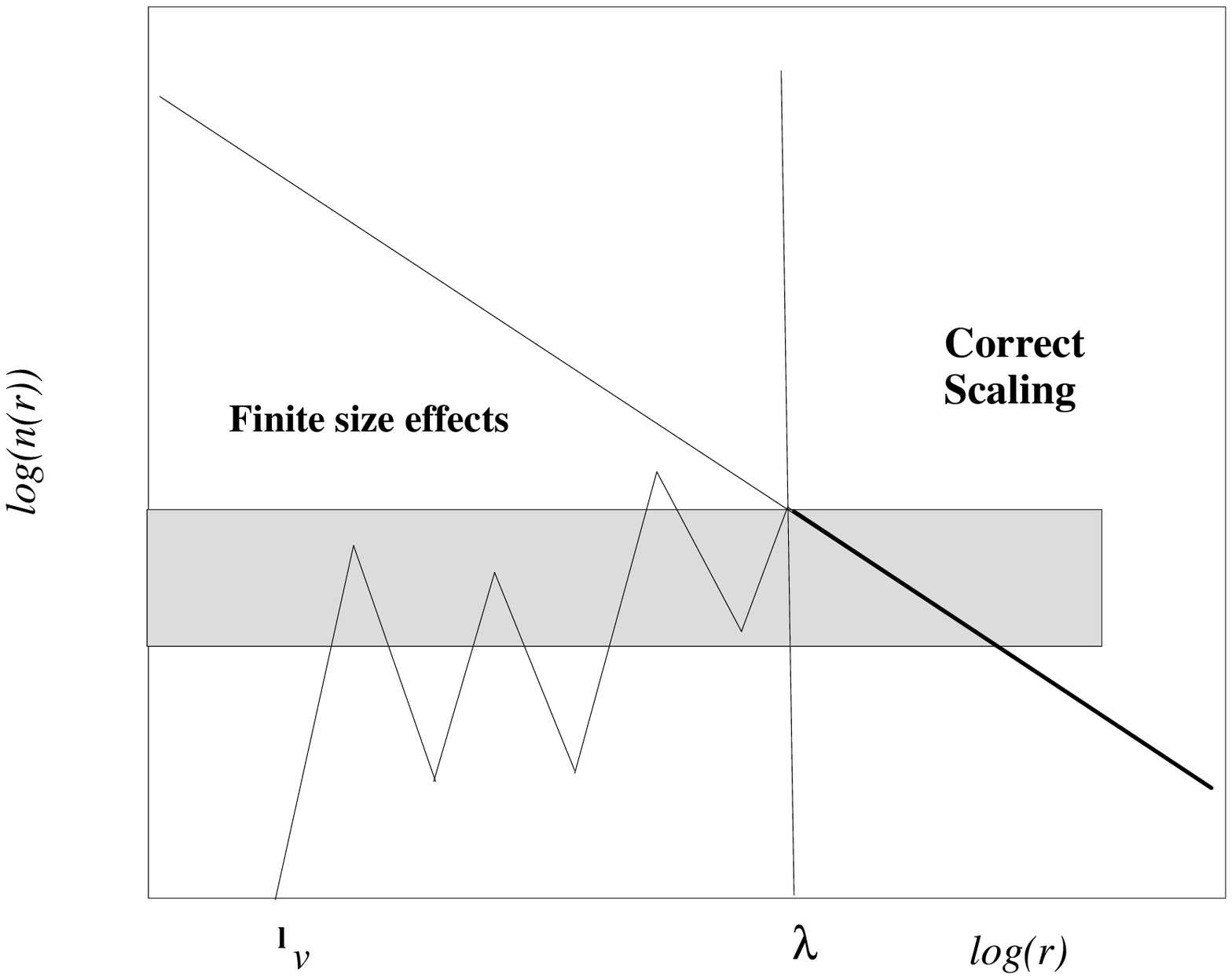}} 
\caption{\label{fig61}  Behavior of the density computed from one
 point, in the case of a fractal
distribution. At small distances
below the average mean separation between neighbor galaxies,
one finds no galaxies. 
Then the number of galaxies starts to grow, but this regime
 is strongly affected by finite size fluctuations. Finally the
 correct scaling region $r \approx \lambda$ is reached. In the
 intermediate region the density can be approximated roughly by a
 constant value.   This leads to an apparent exponent $D \approx
 3$.    This exponent is not real but just due to the size effects.
 } 
 \eef 
For a Poisson sample consisting of $N$ particles inside a volume
 $V$ then the Voronoi volume 
is of the order 
\be
 \label{v1} 
v_v \sim \frac{V}{N} 
\ee 
and $\ell_v \approx v_v^{1/3}$. In the case of homogeneous
 distributions, where the fluctuations have  {\em small amplitude}
 with respect to the average density, one  readily recovers the
 statistical properties of the system at small distances, say, $ r
 \gtapprox 5 \ell_v$. 

The case of fractal distribution is more subtle.
Instead of the Voronoi length we can consider the average distance between 
nearest neighbors, but for clarity we proceed in the following way.
  Eq.\ref{l1}
gives the mass-length relation for a fractal.
In this case, the prefactor $B$ is defined for spherical  samples.
 If we have a sample consisting in a  portion of 
 a sphere characterized by a solid angle
 $\Omega$, we write 
\be 
\label{new1a} 
N(<R) = B R^D \frac{\Omega}{4\pi} \;.
\ee 

In the case of a finite fractal structure, we have to take into account 
the statistical fluctuations. 
We can identify two basic kinds of fluctuations:
the first ones are intrinsic $f(R)$  and are due to
the highly fluctuating nature of fractal distributions
while the second ones, $P(R)$, are Poissonian fluctuations.
Concerning the first ones, 
one has to consider that 
the mass-length relation is a convolution of fluctuations
which are present at all scales.
For example one   encounters, at any scale,  a 
large scale structure  and 
then a huge void: these fluctuations   affect the power law behavior of 
$N(<R)$. We can quantify these effects as a {\it modulating term} 
around the expected average given by Eq.\ref{new1a}. 
Therefore, in the observations from a single point {\it "i"} 
we   have 
\be
\label{cazzz4} 
[N(<R)]_i = B R^D \frac{\Omega}{4\pi} \cdot f_{\Omega}(R, \delta \Omega) \; .
\ee 
In general it is more useful to focus on the behaviour
of a local quantity (as the number of points in shells) rather
than   an integrated one. However for the purpose of the present
discussion the approximation given by Eq.\ref{cazzz4} is rather good.
This equation  shows that the amplitude of $N(<r)$ is related to 
the amplitude of the intrinsic fluctuations and not only 
to the lower cut-off $B$.
The amplitude  
of the modulating term
is small, compared with the expected value of $N(<R)$
\be 
\label{cazzz1}
\sqrt{|f(R)|^2} < B R^D \frac{\Omega}{4\pi} \;.
\ee
 In general this
fluctuating term  depends on the direction of observation $\Omega$ 
and on  the solid angle of the survey $\delta \Omega$
so that $f(R) = f_{\Omega} (R, \delta \Omega)$.  If we have a 
spherical sample we get
\be
\label{cazzz2}
\frac{1}{4 \pi} \sum_{\Omega} 
f_{\Omega}(R, \delta \Omega) =  f_{4 \pi} (R)  \; .
\ee
In general we expect that $f_{\Omega}(R, \delta \Omega) > f_{4 \pi} (R)$, so 
that larger is the solid angle and smaller is the effect this term.
If we perform the ensemble average of this fluctuating term we 
can smooth out its effect and we have then
\be
\label{cazzz3}
\langle f_{4 \pi} (R) \rangle_i = 1
\ee
 where the average $\langle .... \rangle_i$ refers to all the 
 occupied points in the 
 sample. In such a way
 the conditional density, averaged over all the points
 of the sample,  has a single power law behavior. 
We stress that 
according to Eq.\ref{cazzz4} the fluctuations in the 
integrated number of points in a fractal, are proportional
to the number of points itself, rather than to the 
root mean square as in a poissonian distribution.
In general \cite{badii84,smith86,sor96,solis97} 
it is possible to characterize these intrinsic fluctuations
as log-periodic oscillations in the power law behavior.
By performing an ensemble average as in Eq.\ref{cazzz3} 
these oscillations can be smoothed out.
However for the purpose of the present paper,
we limit our discussion to the approximation of Eq.\ref{cazzz4}, 
without entering in more details.

 The second $P(R)$
 term is an additive one, and it takes
 into account spurious finite size fluctuations
is simply due to shot noise. In this case we  have that
\be 
\label{cazzz5}
\sqrt{|P(R)|^2} > B R^D \frac{\Omega}{4\pi} \;\;\; \mbox{if}
 \;\;\; R < \lambda \;
\ee
while $P(R) \approx 0$ for $ R \gtapprox \lambda$.  
The ensemble average is, again,  expected to be 
\be
\label{cazzz7}
\langle P(R) \rangle_i = 0 \; .
\ee
This term becomes negligible if the shot noise fluctuations
are small: for example, if
\be
\label{cazzz8}
[N(<R)]_i  > 10 \sqrt{[N(<R)]_i}    \; .
\ee
From this condition and Eq.\ref{cazzz4} we can have a condition on $\lambda$
(neglecting the effect of $f(R)$):
\be
\label{cazzz9}
\langle \lambda \rangle 
\sim \left(10^2 \frac{4 \pi} {B \Omega} \right)^{\frac{1}{D}}
\ee
The {\em minimal statistical length} $\lambda$ is an explicit
 function of the prefactor $B$ and of the solid angle
 of the survey $\Omega$. This scale is a lower limit for the scaling 
region of the distribution: the effect of intrinsic fluctuations,
described by $f(R)$, which are in general non negligible, 
 can modulate the distance at which 
the scaling region is reached. This length depends also, but weakly, 
on the particular
 morphological features of the sample.  
Therefore it is important to stress that Eq.\ref{cazzz9} 
gives an order of magnitude
for $\langle \lambda \rangle $, where we intend the average value over 
all the possible directions of observations. In different 
directions one can have different values for $\langle \lambda \rangle $,
because of the effect of $f(r)$.

 In the case of real galaxy catalogs we have to consider the
 luminosity selection effects. In a VL sample,
 characterized by an absolute magnitude limit $M_{lim}$,  
   the mass-length relation  Eq.\ref{new1a},  can be
 generalized as 
\be 
\label{new6} 
N(R,M_{lim}) = B R^D \frac{\Omega}{4 \pi} \psi(M_{lim})
 \ee 
where $\psi(M_{lim})$ is the probability that a galaxy has an
 absolute magnitude brighter than $M_{lim}$ 
\be 
\label{new7} 
0 < \psi(M_{lim})  = \frac{\int_{-\infty}^{M_{lim}} \phi(M) dM}
 {\Psi(\infty)} < 1 
\ee 
where $\phi(M)$ is the Schecther 
luminosity function (normalized to unity) and
$\Psi(\infty)$ is the normalizing factor 
\be 
\label{new8} 
\Psi(\infty) = \int_{-\infty}^{M_{min}} \phi(M) dM
\ee 
where $M_{min}$ is the fainter absolute magnitude surveyed in the
 catalog (usually $M_{min} \approx -10 \div -11$).

  It is possible to compute the intrinsic prefactor $B$ from the
 knowledge of the conditional density $\Gamma(r)$ (Eq.\ref{g2} 
 \cite{cp92,slmp97})
computed in the VL samples and normalized for the luminosity factor
 (Eq.\ref{new7}). In the various VL  subsamples of Perseus-Pisces,
 CfA1, and other  redshift surveys  we find
 that 
\be 
\label{new10} 
B \approx 10 \div 15 (\hmp)^{-D}
 \ee
 depending on the parameters of the Schecther function $M^*$ and
 $\delta$.
 From Eq.\ref{cazzz9}, Eq.\ref{new6} and Eq.\ref{new10}  we obtain 
for a typical volume limited sample with $M_{lim} \approx M^*$,
\be
\label{v3} 
\langle \lambda \rangle \approx \frac{(20 \div 60) 
\hmp}{\Omega^{\frac{1}{D}}} \;.
\ee 
 This is the value of the {\em minimal statistical length} that we
  use in what follows.
In  Tab. \ref{tablambda} we report the value of $\lambda$ for several 
redshift surveys. While in the case of CfA1, SSRS1, 
PP, LEDA and ESP we have checked that there is a 
reasonable agreement with this 
prediction,  the CfA2 and SSRS2 
redshift surveys are not sill published and hence in these
cases we can {\it predict} the value of $\lambda$.

\begin{table} \begin{center} 
\begin{tabular}{|c|c|c|} 
\hline         
&       &     \\ 
Survey & $\Omega (sr)$ & $\lambda (\hmp)$  \\ 
   &    &     \\ 
\hline 
CfA1            & 1.8      & 15 \\  
CfA2 (North)& 1.3      & 20  \\        
SSRS1          & 1.75    & 15  \\    
SSRS2          & 1.13    & 20 \\  
PP                & 1         & 40  \\
LEDA            & 2 $\pi$& 10 \\      
IRAS            & 2 $\pi$ & 15 \\        
ESP             &   0.006  & 300  \\        
\hline
\end{tabular} 
\caption{The {\em minimal statistical length 
\label{tablambda}} $\lambda$ for several redshift surveys }
\end{center} \end{table}

Given the previous discussion, we can now describe in a very simple 
way the behavior of $[N(<r)]_i$, i.e. the mass length relation 
measured from a generic point {\it "i"}. Given a sample 
 with solid angle $\Omega$, we can approximate 
 the effect of the intrinsic and shot noise fluctuations in the following way:
 \be 
 \label{equ22}
(N(<r))_i =   B_1 r^3 \; \; \; \mbox{if} 
\; \;  r \ltapprox \lambda
\ee
i.e. the density is constant up to $\lambda$, while 
\be 
 \label{equ23}
(N(<r))_i  =  B  r^{D} \; \; \; \mbox{if} 
\; \;  r \gtapprox \lambda
\ee
so that by the condition of continuity at $\lambda$ we have 
\be 
 \label{equ24}
B_1= \frac{B}{ \lambda^{3-D}}
\ee
This simple approximation is very useful in the following discussion, 
especially for the number counts.

To clarify the effects of the spatial inhomogeneities and finite
effects we have studied the behavior of  the galaxy radial
 (number)
density in the VL samples, i.e. the behavior of (using Eq.\ref{new6}
and Eq.\ref{cazzz4})
\be 
\label{g2b} 
n_{VL}(r) = \frac{N(<r)}{V(r)} = \frac{3}{4\pi} B r^{D-3} \psi(M_{lim}) 
\cdot f_{\Omega}(r, \delta \Omega)
\ee  
 One expects that, if the distribution is homogeneous the density
 is  constant,  while if it is fractal it decays as power law.

 When one computes the conditional average density,
 one
 indeed performs an average over all the points of the survey. In
 particular, as we have already discussed,
 we limit our analysis to a size defined by the radius of the
 maximum sphere fully contained in the sample volume, and  we do not 
 make 
  use  any   treatment of the sample
 boundaries.
 On the contrary Eq.\ref{g2b} is computed only from a single  point,
 the origin. This allows us to extend the study of the spatial
 distribution up to very deep scales. 
 The price to pay is that this method is strongly affected by 
 statistical fluctuations and finite size effects.

The effect of the 
finite size spurious fluctuations for a fractal distribution
 is shown Fig.\ref{fig61}:  at small distances one finds  almost no
 galaxies because we are below {\it the average  separation between 
 neighbor galaxies} $\ell$. Then the number
 of galaxies starts to grow, but this regime is strongly affected
 by finite size fluctuations. Finally the correct scaling region $r
 \approx \lambda$ is reached. This means, for example, that if one
 has a fractal distribution, there is first a raise of the
 density, due to finite size effects and characterized by strong 
 fluctuations, because no galaxies are present before a certain
 characteristic scale. Once one enters in the correct scaling
 regime for a fractal the  density starts to decay as a power law.
 So in this regime of raise and fall with strong fluctuations there
 is a region in which  the density can be approximated roughly
 by a  constant value.  This leads to an apparent exponent $D
 \approx 3$, so that the integrated number grows as   $N(<r) \sim
 r^3$.  This exponent is therefore not a real one 
 but just due to  
 finite size fluctuations.
Of course, depending on the survey orientation in the sky, one can get an
exponent larger or smaller  than $3$, but in general this is 
the more frequent situation (see below).
 Only when a well defined statistical
 scaling regime has been reached, i.e. for $r > \lambda$,  one can
 find the genuine scaling properties of the   structure, otherwise
 the behavior is completely  dominated by spurious finite size
 effects (for seek of clarity in this discussion we do not
 consider the effect of $f(r)$ in Eq.\ref{cazzz4})

The question of the difference between the integration from the
 origin (radial density) 
  and correlation properties  averaged over all points lead
 us to a subtle problem of {\em asymmetric fluctuations} in a
 fractal structure. From our discussion, exemplified by
 Fig.\ref{fig61}, the region between $\ell$ and $\lambda$ 
 corresponds to an underdensity with respect to the real one.
 However we have also showed that for the full correlation averaged
 over all the points, as measured by $\Gamma(r)$,
 the correct scaling is recovered
 at distances appreciably smaller than $\lambda$. This means that
 in some points one should observe an overdensity between $\ell$
 and $\lambda$. However, given the intrinsic asymmetry between
 filled and empty regions in a fractal, only very few points 
 show the overdensity (a fractal structure is asymptotically
 dominated by voids). These few points nevertheless  have, indeed,  an
 important effect on the average values of the correlations. This
 means that, in practice, a typical points shows an underdensity up
 to $\lambda$ as shown in Fig.\ref{fig61}. The full averages instead
 converge at much shorter distances. This discussion shows the
{\it  peculiar and asymmetric nature of finite size fluctuations in
 fractals} as compared to the symmetric Poissonian case.  For
 homogeneous distribution\cite{slgmp96}
the situation is in fact quite different. Below the Voronoi length
 $\ell_v$ there are finite size fluctuations, but for distances $r
 \gtapprox (2\div 4) \ell_v$ the correct scaling regime is readily
 found for the density.
 In this case the finite size effects do not affect too much the
 properties of the system because a Poisson distribution is
 characterized by {\em small amplitude fluctuations}. 

As an example, we show the behaviour of the radial density in the Perseus-Pisces 
redshift survey (PPRS).
We have computed the $n(r)$ in the various VL samples 
of PPRS, 
and we show the results   in Fig.\ref{fig62}.
 In the less deeper VL samples (VL70, VL90)   the
 density does not show any smooth behavior because in this case the
 finite size effects dominate the behavior as the distances involved are 
 $r < \lambda $ (Eq.\ref{v3}). At about  the same scales we
  find a very well defined power law behavior by the
 correlation  function analysis.
 In the deeper VL samples (VL110, VL130)  a smooth
 behavior is reached for distances larger 
than the scaling distance ($\Omega =0.9 \, sr$)
 $r \approx \lambda \sim 50 h^{-1}Mpc$. The fractal dimension is
  $D \approx 2$ as one measures by 
  the correlation 
 function.     
\bef 
\epsfxsize 4cm 
\centerline{\epsfbox{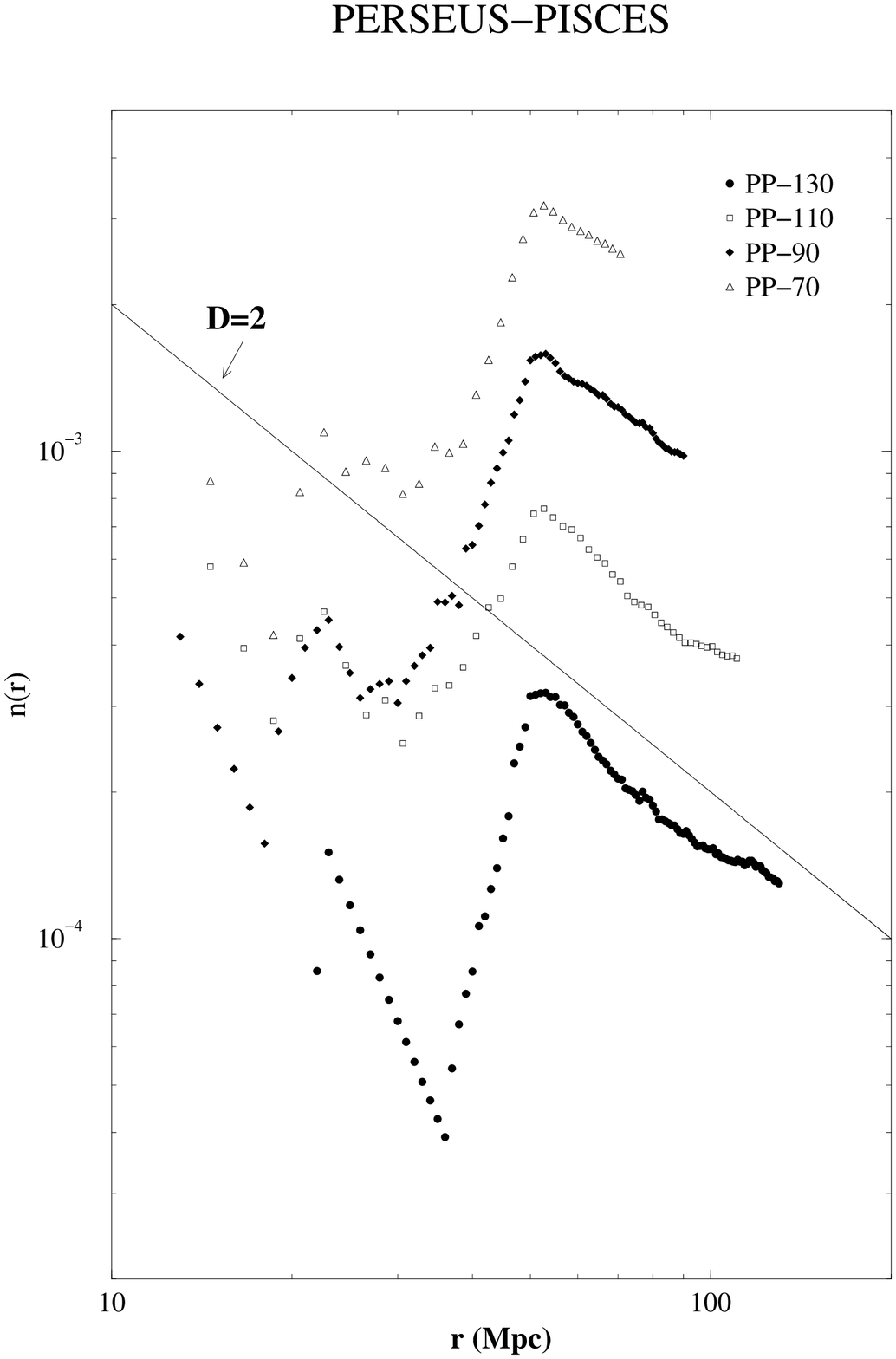}} 
\caption{ The spatial density $n(r)$   computed in the VL sample
 cut at $70, 90, 110, 130  h^{-1}Mpc$ . In the case of VL70 and VL90  the
 density is dominated by large fluctuations and it has not reached
 the scaling regime. In the samples  VL110   and VL130 the density is
 dominated by large fluctuations only at small distances, while at
 larger distances, after the Perseus Pisces chain at $50
 h^{-1}Mpc$, a very  well defined power law behavior is shown, with
 the same exponent of  the correlation function  (i.e. $D=2$)
\label{fig62}}
\eef
For relatively small volumes it is possible to
 recover the correct scaling behavior for scales of order of
 $\ell$ (instead of $ \sim 10 \ell$) by averaging over several
 samples or, as it happens in real cases, over several points of
 the same sample when this is possible. Indeed, when we compute the
 correlation function we perform an average over all the points of
 the system even if the VL sample is not deep enough to satisfy the
 condition expressed by Eq.\ref{v3}. In this case the lower
 cut-off  introduces a limit in the sample statistics \cite{slmp97}.   

The case of LCRS and ESP are discussed in detail 
in Sylos Labini \etal (1997) and the
results are shown in Fig.\ref{fig82}

\subsection{Pencil beams}
\label{radialpencil}

Deep "pencil beams" surveys 
cover in general very narrow angular size ($ \sim 1^{\circ}$)
and extend to very deep depths ($ z\gtapprox 0.2$). 
These narrow shots through deep space provide
a confirmation of strong inhomogeneities in the galaxy distribution.

One of the most discussed results obtained from pencil-beams surveys 
has been the claimed detection of a typical 
scale in the distribution of galaxy structures, corresponding to a
characteristic separation of $128 \hmp$ \cite{bro90}. 
However in the last five years several other surveys, in different
regions of the sky, do not find {\it any} evidence for such a periodicity.
In particular, Bellanger \& De Lapparent \cite{bd95}, by analyzing 
a sample of 353 galaxies in the redshift interval 
$0.1 \ltapprox z \ltapprox 0.5$, concluded that 
these new data contain {\it no evidence for a periodic signal} on a scale 
of $128 \hmp$. Moreover they argue that the low sampling rate of 
\cite{bro90} is insufficient for mapping the detailed large-scale structure
and it is the real origin of the apparent periodicity.

On the other hand, Willmer \etal (1994)
 detect four of the five nearest peaks 
of the galaxies detected by \cite{bro90}, because their survey 
is contiguous to that of \cite{bro90}, in the sky region near the north 
Galactic pole. Moreover Ettori \etal (1996)
in a survey oriented in three small regions around the South Galactic Pole
do not find any statistically periodic signal distinguishable 
from noise.  Finally 
Cohen \etal \cite{coh96} by analyzing a sample of 140 objects up to $z \sim 0.8$, 
find that there is no evidence for periodicity 
in the peak redshifts.

In a pencil-beam survey one can study 
the behavior of the {\it linear density}
along a tiny but very long cylinder. 
The observed galaxy distribution corresponds therefore to the 
intersection of the full three dimensional galaxy distribution with one
dimensional cylinder \cite{cp92}. In this case, from the law of
codimension additivity \cite{man83,cp92}, one
obtains that the fractal dimension of the intersection is given by
\be
\label{pb1}
D_I = D + d_{pc} -d \approx D+1-3 \approx 0
\ee
where $D$ is the galaxy distribution fractal dimension, 
embedded in a $d=3$ Euclidean space, and $d_{pc} =1$ is the 
dimension of the pencil beam survey. 
This means that the set of points visible in a 
randomly oriented cylinder has dimension $D_I \approx 0$.
In such a situation the power law behavior is no longer present and the 
data should show a chaotic, featureless nature strongly dependent
on the beam orientation. If the galaxy distribution   
becomes instead homogeneous above 
some length, shorter than the pencil beam depth, 
one has the regular situation $D_I = 1$ 
and a well defined density
must be observed. 

We  stress that in {\it any} of the available
pencil beams surveys, one can detect   tendency towards 
an homogeneous distribution. Rather, all these surveys show a
very fluctuating signal, characterized by the presence of galaxy structures. 
Some authors \cite{sch96} claim to detect the {\it end greatness} (i.e.
that the galaxy structures in the deep pencil beams are not so different
from those seen in nearby sample - as the Great Wall), or that
\cite{bd95} the dimension of voids does not scale with sample size, by the 
visual inspection of these surveys. However one should consider 
in these morphological analyses, that one is just looking at a convolution of the 
survey geometry, which in general are characterized by very narrow solid angles, 
 and large scale structures, and that in such a situation,
a part from very favorable cases, one may detect portions of 
galaxy structures (or voids).

Finally we note that if the periodicity would be present, 
the amplitude of the different peaks is very different from each other, and 
an eventual transition to homogeneity in a periodic lattice, should be, for
example, ten times the lattice parameter, i.e. $\gtapprox 1000 \hmp$ !


\section{Number Counts}

  The most complete information about galaxy distribution comes
 from the full three dimensional samples, while  the angular
 catalogs have a poorer qualitative information, even if usually
 they contain  a much larger number of galaxies. However, one of
 the most important tasks in observational astrophysics,  is the
 determination of the  $\log N-\log S$ relation for different kind
 of objects: galaxies in the  various spectral  band (from
 ultraviolet to infrared), radio-galaxies, Quasars, X-ray sources
 and $\gamma$-ray bursts. This relation gives the number $N(S)$
 (integral or differential)  of objects, for unit  solid angle,
 with {\it apparent flux} (larger than a certain limit) $S$. The
 determination of such a quantity avoids the measurements of the
 distance, which is always a very complex task. However we show in
 the following that the behavior of the $\log N-\log S$ is
 strongly biased by some statistical finite size effects 
due to small scale fluctuations.

The counts of galaxies as  a function of the apparent magnitude are
determined from the Earth only, and hence it is not possible
to make an average over different observers. As we have 
already discussed in the previous section, this kind of measurement 
is affected by intrinsic fluctuations that are not smoothed out at any scale.
Moreover at small scale there are finite size effects which may
seriously perturb the behavior of the observed counts. Following the 
simple argument about the behaviour of the radial density 
we have presented in the previous section, we consider here the 
problem of the galaxy-number counts.

 We present in this section a
 new interpretation of this basic relation at the light of the 
 highly inhomogeneity nature of galaxy distribution, and we show
 its compatibility with the behavior of  counts of  galaxies in
 different frequency bands, radiogalaxies, Quasars and X-ray
 sources. {\it Our  conclusion is that the counts of all
 these different kind of objects are compatible with a fractal
 distribution of visible matter up  to the deepest observable
 scale.}

\subsection{Galaxy number counts data} 
\label{countsdata}

  Historically \cite{hu26,pee93} the
 oldest type of data about galaxy distribution is given by the
 relation between the number of observed galaxies $N(>S)$  and
 their apparent brightness $S$. It is easy to show that,
under very general assumptions one gets 
(see Sec.\ref{countsbasic})
 \begin{equation} 
\label{eq1} 
N(>S)  \sim  S^{-\frac{D}{2}}
 \end{equation} 
where $D$  is the fractal dimension of the galaxy
 distribution. Usually this relation is written in terms of the
 apparent magnitude $m$ ($S \sim 10^{-0.4 m}$ - 
note that bright galaxies
 correspond to small $m$). In terms of $m$, Eq.\ref{eq1} becomes
\be 
\label{nn1} 
\log N(<m)   \sim \alpha m 
\ee 
 with $\alpha =
 D/5$ \cite{bslmp94,pee93}. Note that $\alpha$ is the
 coefficient of the exponential behavior of Eq.\ref{nn1} and we
 call it "exponent" even though  it  should not be confused
 with the exponents of power law behaviors. In Fig.\ref{fig85} 
\bef 
\epsfxsize8cm
\centerline{\epsfbox{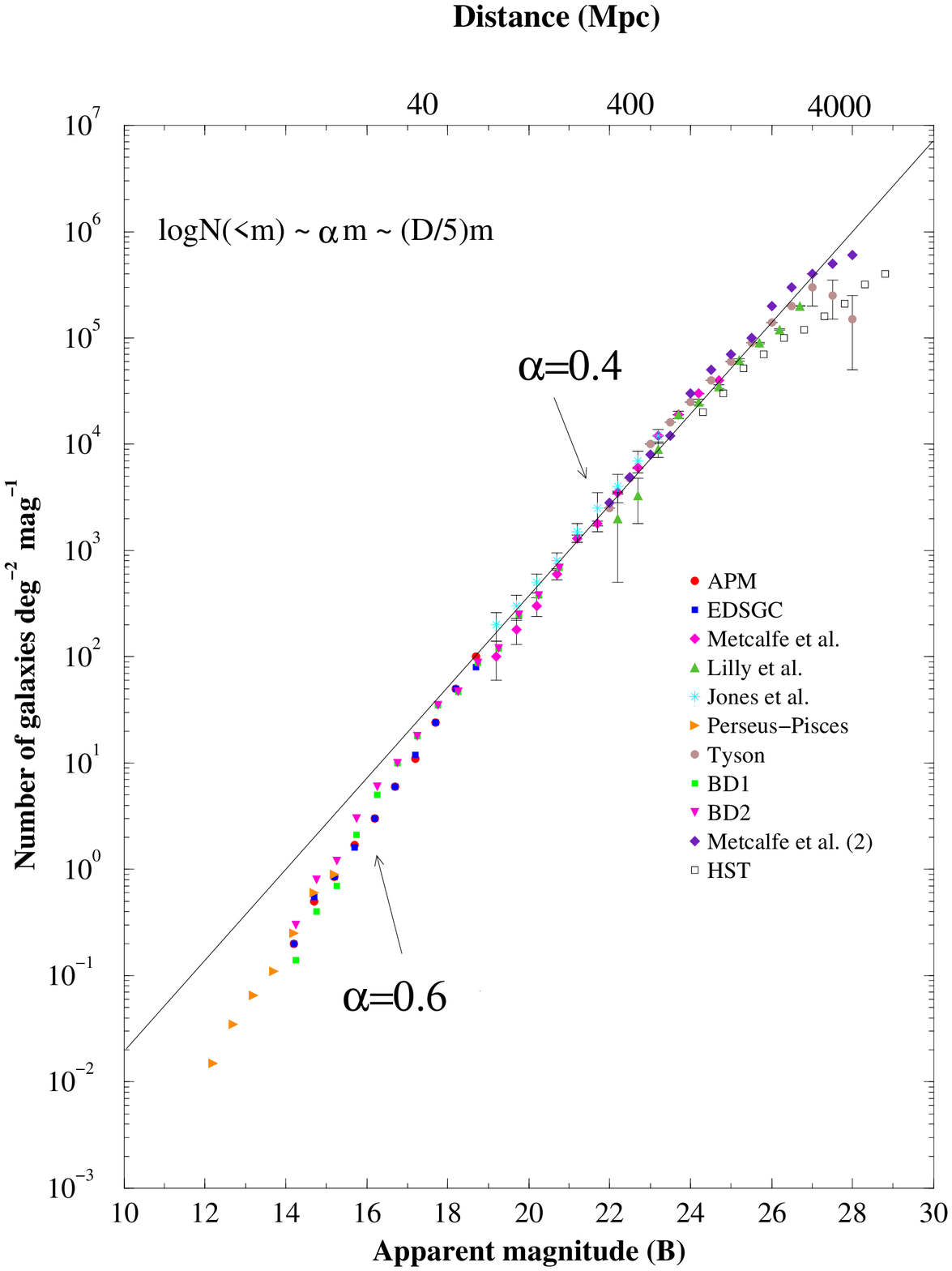}}
\caption{ \label{fig85} The galaxy number
 counts in the $B$-band, from several surveys. 
In the range $12 \ltapprox  m \ltapprox
 19$ the counts show an exponent $\alpha \approx 0.6 \pm 0.1$,
 while in the
 range $19 \ltapprox m \ltapprox 28$ the exponent is $\alpha 
 \approx 0.4$. The solid line is computed from the determination
of the amplitude of the conditional density at small scale, 
the fractal dimension  $D=2.2$, and from the knowledge
of the luminosity function.
The distance is computed for a galaxy with $M=-16$ 
and we have used $H_0=75 km sec^ {-1} Mpc^{-1}$.}  
\eef
 we have collected all the recent observations
 of $N(<m)$ versus $m$ in the
 $B$-spectral-band $m_B$ 
 \cite{ms91,ty88,lc91,jf91,dr94,cg90,co91}.
 At bright and intermediate magnitudes
 ($\:12 \ltapprox m_B \ltapprox 18$), corresponding to small
 redshift ($\:z<0.2$),  one obtains $\:\alpha \approx 0.6$, while
 from $m_B \sim 19$ up to $m_B \sim 28$ the counts are well fitted
 by a smaller exponent with $\alpha \approx  0.4$.  The usual
 interpretation \cite{pee93,yo93,be92,yt88,yp95}  is that
 $\alpha \approx  0.6$ corresponds to $D \approx 3$ consistent with
 homogeneity, while at large scales galaxy evolution and space time
 expansion effects are invoked to  explain the lower value $\alpha 
 \approx 0.4$. On the basis of the previous discussion of the VL
 samples,   this interpretation is untenable. In fact,
 we know for sure that, at least  up to 
$\sim 150  \hmp $ there
 are fractal correlations, as we have discussed 
in the previous sections,
 so one would eventually expect the opposite
 behavior. Namely  small value of $\alpha \approx 0.4$ (consistent
 with $D \approx 2$) at small scales followed by a crossover to an
 eventual homogeneous distribution at large scales ($\alpha 
 \approx 0.6$ and $D  \approx 3$).

The situation is therefore
 similar in the different spectral bands. The puzzling
 behavior of the GNC represents an important apparent 
contradiction we
 find in the data analysis.   We argue here  that this  apparently
 contradictory experimental situation can be fully understood on 
 the light of the small scale effects in the space distribution of
 galaxies. For example a fractal distribution
 is non analytic in
 every occupied point: it is not possible to  define a meaningful
 average density because  we are dealing with intrinsic
 fluctuations which grow with  as the scale of the system itself.
 This situation is qualitatively different from an homogeneous 
 picture, in which a well defined density exists, 
and the
 fluctuations represent only  small amplitude perturbations. The
 nature of the fluctuations in these two cases is 
completely
 different, and for fractals the fluctuations themselves define all
 the statistical  properties of the distribution.
 This concept has 
 dramatic consequences in the following discussion as well as in
 the  determination of various observable quantities, such as the
 amplitude of the two point  angular  
correlation function.   It is
 worth to notice that the
  small scale effects are usually neglected in
 the study of fractal structures because one can generate large
 enough (artificial) 
structures to avoid these problems. In Astrophysics the
 data are instead intrinsically 
 limited and  
 a detailed analysis of finite size effects is very important. We
 discuss     the problems of finite size effects in the
 determination of the asymptotic properties of fractal
 distributions,  considering explicitly the problems induced by the
 lower cut-off.

\subsection{Galaxy counts: basic relations} 
\label{countsbasic}

We briefly introduce some basic relations which are 
 useful later. We can start by computing
the expected GNC in
 the simplest case of a magnitude limited (ML) sample. A ML sample
 is obtained by measuring all the galaxies with apparent magnitude
 brighter than a certain limit $m_{lim}$. In this case we have (for
 $m < m_{lim}$) 
\be 
\label{q3} 
N(<m) = 
B \Phi(\infty) 10^{\frac{D}{5}m} 
\ee where 
\be 
\label{q3a} 
\Phi(\infty) =
 \int_{-\infty}^{\infty} \phi(M) 10^{-\frac{D}{5}(M+25)} dM \;.
\ee  
We
 consider now the case of a volume limited (VL) sample. A  VL 
 sample consists of  every galaxy in the volume which is more luminous
 than a certain limit, so that in such a  
sample there is no
 incompleteness for an observational luminosity selection effect
 \cite{dp83,cp92}. Such a sample is 
defined by a certain
 maximum distance $R$ and an absolute magnitude limit given by: 
\be
 \label{q3b}
 M_{lim}=m_{lim}-5\log_{10}R -25 - A(z)
\ee
 ($m_{lim}$ is the
 survey apparent magnitude limit). By performing the calculations
 for the number-magnitude relation,   we obtain 
\be
\label{q4b}
 N(<m)= A(m) \cdot 10^{\frac{D}{5}m} + C(m) 
\ee 
where $A(m)$ is 
\be
 \label{q4c} 
A(m) =B \int_{M(m)}^{M_{lim}}\phi(M)
 10^{-\frac{D}{5}(M+25)}dM
\ee 
and $M(m)$ is given by $M(m) = m - 5 \log(R) -25$,
  and it is a function of $m$.
 The second term is 
\be 
\label{q5e}
C(m)= BR^{D}\int_{-\infty}^{M(m)} \phi(M) dM \;.
\ee 
This term, as $A(m)$, 
depends on the VL sample considered. We assume a luminosity
 function with a Schecther shape. 
 For $M(m) \gtapprox M^{*}$ we have that $C(m)
 \approx 0$, and $A(m)$ is nearly constant with $m$. This happens
 in particular for the deeper VL samples for which $M_{lim}\sim
 M^*$. For the less deeper VL ($M_{lim} > M^*$) samples these terms
 can be considered as a deviation from a power law behavior only
 for $m \rightarrow m_{lim}$. 
 If one has $\phi(M) = \delta(M-M_0)$ then it is simple to show that
 $\log(N(<m)) \sim (D/5)m$ also in each  VL sample.

\subsection{Galaxy counts in redshift surveys}
\label{countsred}

We have studied   the GNC in 
the Perseus-Pisces
 redshift  survey \cite{hg88} in order 
 to clarify the role of spatial inhomogeneities and finite
 size effects. In the previous sections 
 we have analyzed the spatial
 properties of galaxy distribution in this sample by measuring
the conditional (average) density and the radial density.
Let us briefly summarize our main results.

 When one
 computes the conditional average density, one indeed
 performs an average over all the points of the survey.
 On the contrary the radial density is computed only from
 a single  point, the origin. 
This allows us to extend the study of
 the spatial distribution up to very deep scales: 
the price to pay is that this method
 is strongly affected by  statistical fluctuations and finite size
 effects. Analogously, when one computes $N(<m)$,  one does not
 perform an average but  just counts the points from 
the origin. As
 in the case of the radial density
 $n(r)$ also $N(<m)$ is strongly affected by
 statistical fluctuations  due to finite size effects.
 as well as intrinsic oscillations that are not
 smoothed out.
  We are now
 able to 
 clarify how the behavior of $N(<m)$, and in particular its
 exponent,  are influenced  by these effects.

We show in
 Fig.\ref{fig86}, 
\bef 
\epsfxsize 6cm
\centerline{\epsfbox{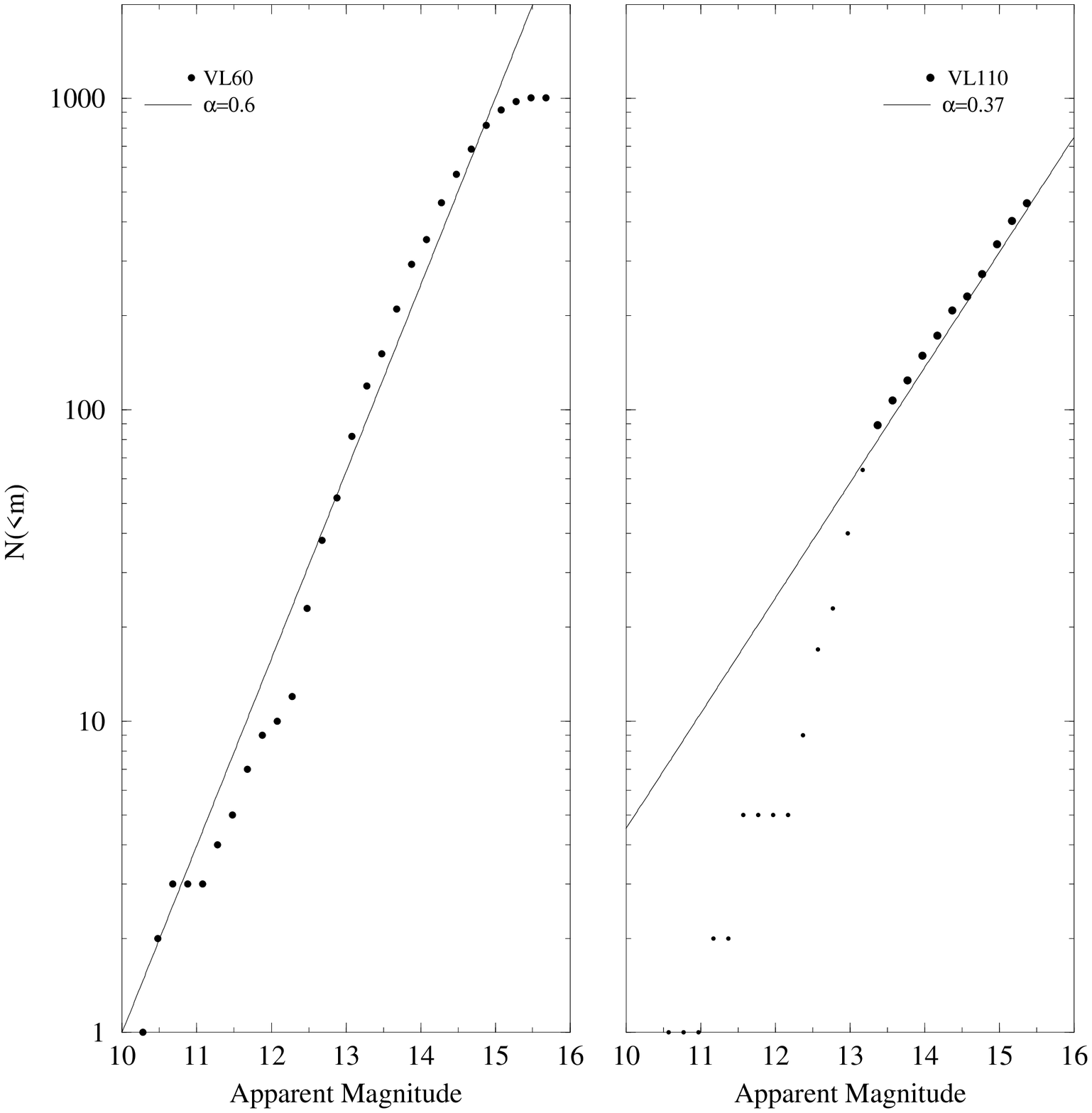}}
\caption{\label{fig86}
$(a)$
 The Number counts $N(<m)$ for the VL sample VL60. The 
 slope is $\alpha \approx 0.6$. In this case occurs a flattening
 for $m \rightarrow m_{lim}$. 
$(b)$ The Number
 counts $N(<m)$ for the VL sample VL110. The slope is
 $\alpha \approx 0.4$, a part from the initial fast growth 
due to weak statistics. This
 behavior corresponds to a well defined define power law behavior of
 the density with exponent $D \approx  5 \alpha \approx 2$.
 } 
\eef 
and 
Fig.\ref{fig87} 
\bef 
\epsfxsize 5cm
\centerline{\epsfbox{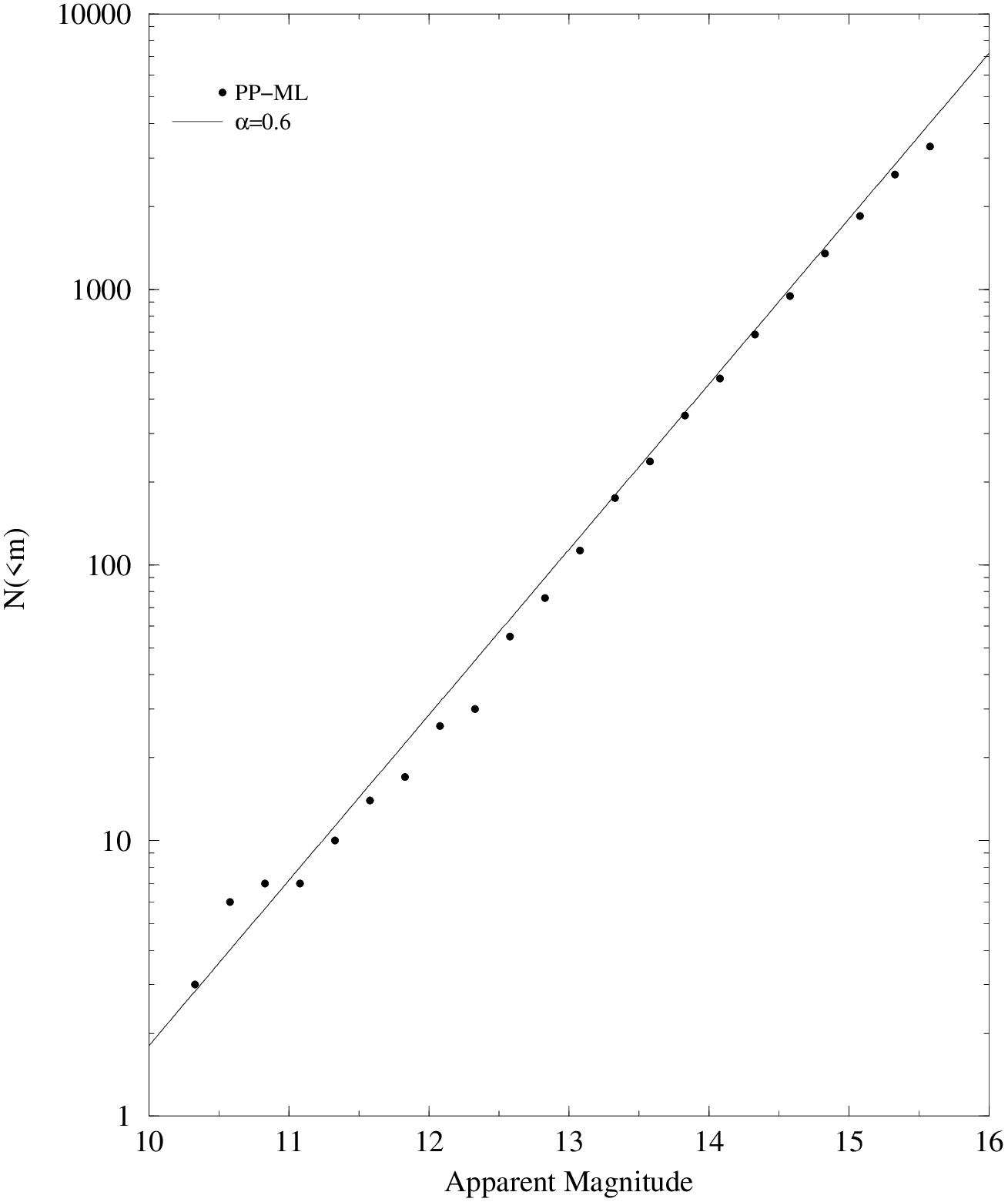}}
 \caption{\label{fig87}
 The Number counts $N(<m)$ for the whole magnitude
 limit sample. The slope is $\alpha \approx 0.6$ and it is clearly
 associated only to fluctuations in
 the spatial distribution rather
 than to a real homogeneity in space. } 
\eef 
the behavior of
 $N(<m)$ respectively for the various VL samples  
 and for the whole
 magnitude
 limit sample. In  VL60 there
 are very strong inhomogeneities in the behavior of $n(r)$ (see
 Fig.\ref{fig62}) and
 these are associated with a slope $\alpha \approx 0.6$ for the
 GNC. (The flattening for $m \rightarrow m_{lim}$ is just due to a
 luminosity selection effect  that is  explained in Sec.\ref{countsbasic}). 
 For VL110 the behavior of the density is much more
 regular and smooth, so that it shows  indeed a clear power law
 behavior. Correspondingly the behavior of $N(<m)$ is well fitted
 by  $\alpha \approx 0.4$. Finally the whole magnitude limit sample
 is again described by an exponent $\alpha \approx 0.6$. 

 We have
 now enough elements to describe the behavior of the  GNC. The
 first point is that the exponent of the GNC is strongly related to
 the space distribution. Indeed, what has never been taken into
 account   before is  the role of finite size effects \cite{slgmp96}.
 The behavior of the GNC is determined by 
 a convolution of the space density
 and the luminosity function, and the space density enters
 in the GNC as an integrated quantity.  The problem is to consider
 the correct space density in the interpretation of   data analysis. 
 In fact, if the density has
 a very fluctuating behavior  
in a certain region of length scale, 
 as in  the case shown in Fig.\ref{fig62}, its integral  over this
 range of length scales is almost equivalent to a flat one. This
 can be seen also in Fig.\ref{fig61}:  at small 
distances one finds 
 almost no galaxies because one is below the mean  minimum separation
 between neighbor galaxies. Then
 the number of galaxies starts to grow, but this regime is strongly
 affected by finite size fluctuations. Finally the correct scaling
 region $r \approx \lambda$ is reached. This
 means for example that
 if one has a fractal distribution, there is first a raise of
 the density, due to finite size effects and characterized by
 strong  fluctuations, because no galaxies are present before a
 certain characteristic scale. Once one enters in the correct
 scaling regime for a fractal the  density becomes to decay as
  a power law. So in this regime of raise and fall with strong
 fluctuations there is  a region in which  the density can be
 approximated roughly by a  constant value.  This leads to an
 apparent exponent $D \approx 3$, so that the integrated number
 grows as   $N(<r) \sim r^3$ and  it is associated in terms of GNC,
 to $\alpha  \approx 0.6$ (Fig.\ref{fig88}).
\bef 
\epsfxsize 7cm
\centerline{\epsfbox{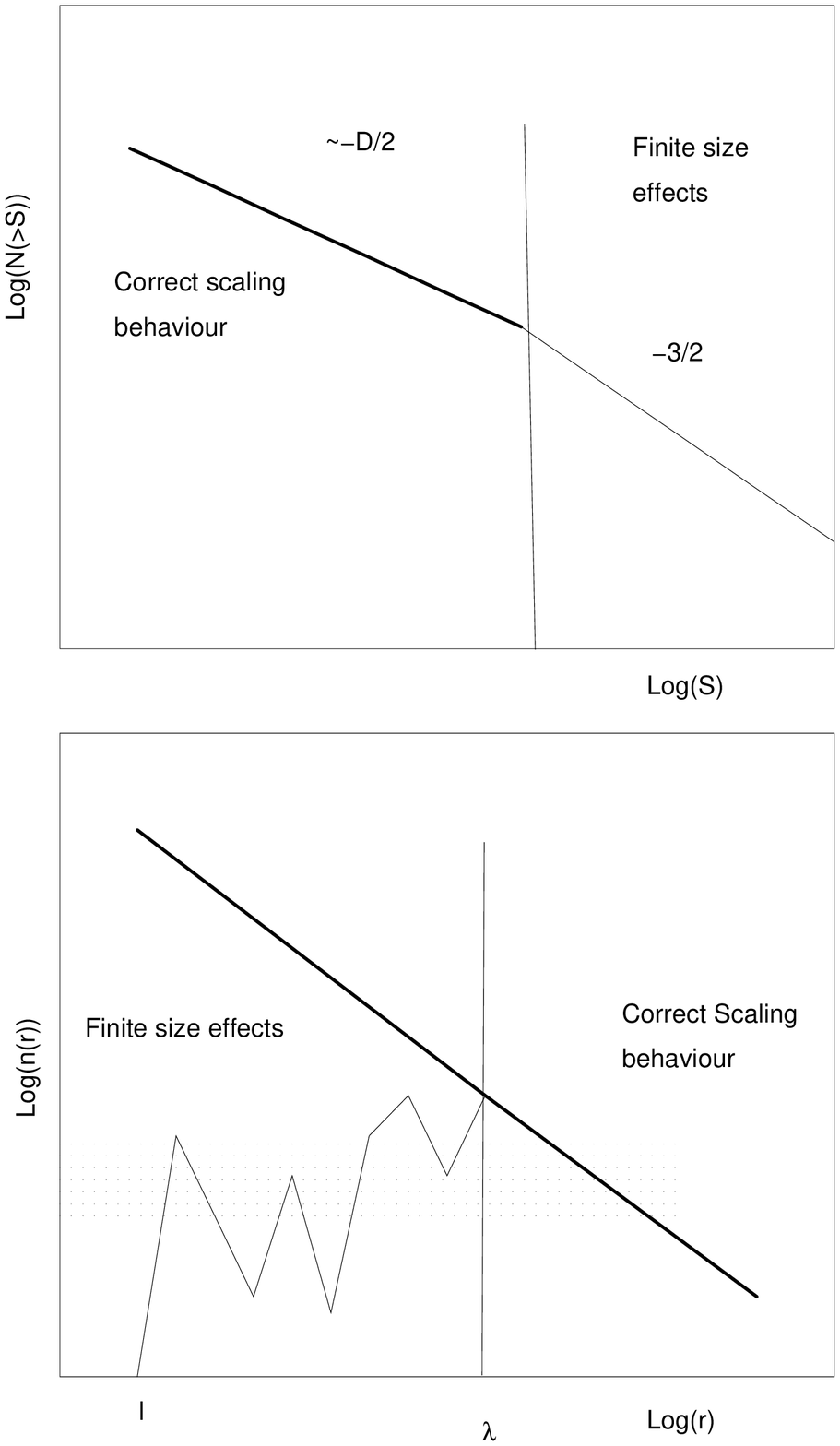}}
 \caption{\label{fig88}
 The number counts $N(<m)$ together with the behavior of 
the space density.
At small scale the density is characterized by
having strong fluctuations which lead to a 
slope $\alpha \approx 0.6$. This is clearly
 associated only to fluctuations in the 
spatial distribution rather
 than to a real homogeneity in space. At larger scales (faint end)
the correct scaling behavior is recovered and $\alpha = D/5$. } 
\eef 
This exponent is therefore not a real one but
 just due to finite size fluctuations. Only when a well defined
 statistical scaling regime has been reached, i.e. for $r >
 \lambda$,  one can find the genuine scaling properties of the  
 structure, otherwise the behavior is completely  dominated by
 spurious finite size effects.  
 In the VL samples where $n(r)$ scales
 with the asymptotic properties (Fig.\ref{fig86}) the GNC grows also
 with the right exponent ($\alpha=D/5$).

 If we now consider instead
 the behavior of the GNC in the whole magnitude limit 
 (hereafter ML) survey, we
 find that the exponent is $\alpha  \approx 0.6$ (Fig.\ref{fig87}). 
 This behavior can be understood by 
 considering that at small
 distances, well inside the distance $\lambda$ defined by
 Eq.\ref{v3}, the number of galaxies present in  the sample is
 large because there are  galaxies of all magnitudes. Hence the
 majority of galaxies correspond to small distances ($r < \lambda$)
 and the distribution has not reached the scaling regime  in
 which the  statistical self-averaging   properties of the system
 are present. For this reason in the ML sample the finite size
 fluctuations dominate completely the behavior of the GNC.
 Therefore this behavior  in the ML sample is associated with
 spurious finite size effects rather than to real homogeneity.
 We discuss in a more quantitative way 
 the behavior in ML surveys later.

\subsection{Test on finite size effects: the average $N(<m)$}
\label{countstest}

To prove that the behavior found in Fig.\ref{fig87}, i.e. that the
 exponent $\alpha \approx 0.6$ is connected to large fluctuations
 due to finite size effects in the space distribution and not to 
 real homogeneity, we have done the 
following test. We have adopted
 the same procedure used for the
 computation of the correlation
 function, i.e. we make an average
 for $N(<m)$ from all the points of the sample rather than counting
 it from the origin only.  

To this aim we have considered a VL
 sample with $N$ galaxies and we have built $N$ independent
 flux-limited surveys in the following way.
 We consider each galaxy
 in the sample as the observer, and for each observer we have
 computed the apparent magnitudes of all the other galaxies. To
 avoid any selection effect we consider only the galaxies  which
 lie inside a well defined volume around the observer. This volume
 is defined by the maximum sphere fully contained in the sample
 volume with the observer as a center.  

Moreover we have another
 selection effect due to the fact that our VL sample has been built
 from a ML survey done with respect to the origin. To avoid this
 incompleteness we have assigned to each galaxy a constant
 magnitude $M$. In fact, our aim is to show that 
the inhomogeneity
 in the space distribution plays the fundamental role that
 determines the shape of the $N(<m)$ relation, and the functional
 form of the luminosity function enters in Eq.\ref{q4c} only as an
 overall normalizing factor.  

Once we have computed $N_i(<m)$ from
 all the points $i=1,..,N$ we then compute the average. We show in
Fig.\ref{fig89} 
\bef 
\epsfxsize 6cm
\epsfysize 6cm
\centerline{\epsfbox{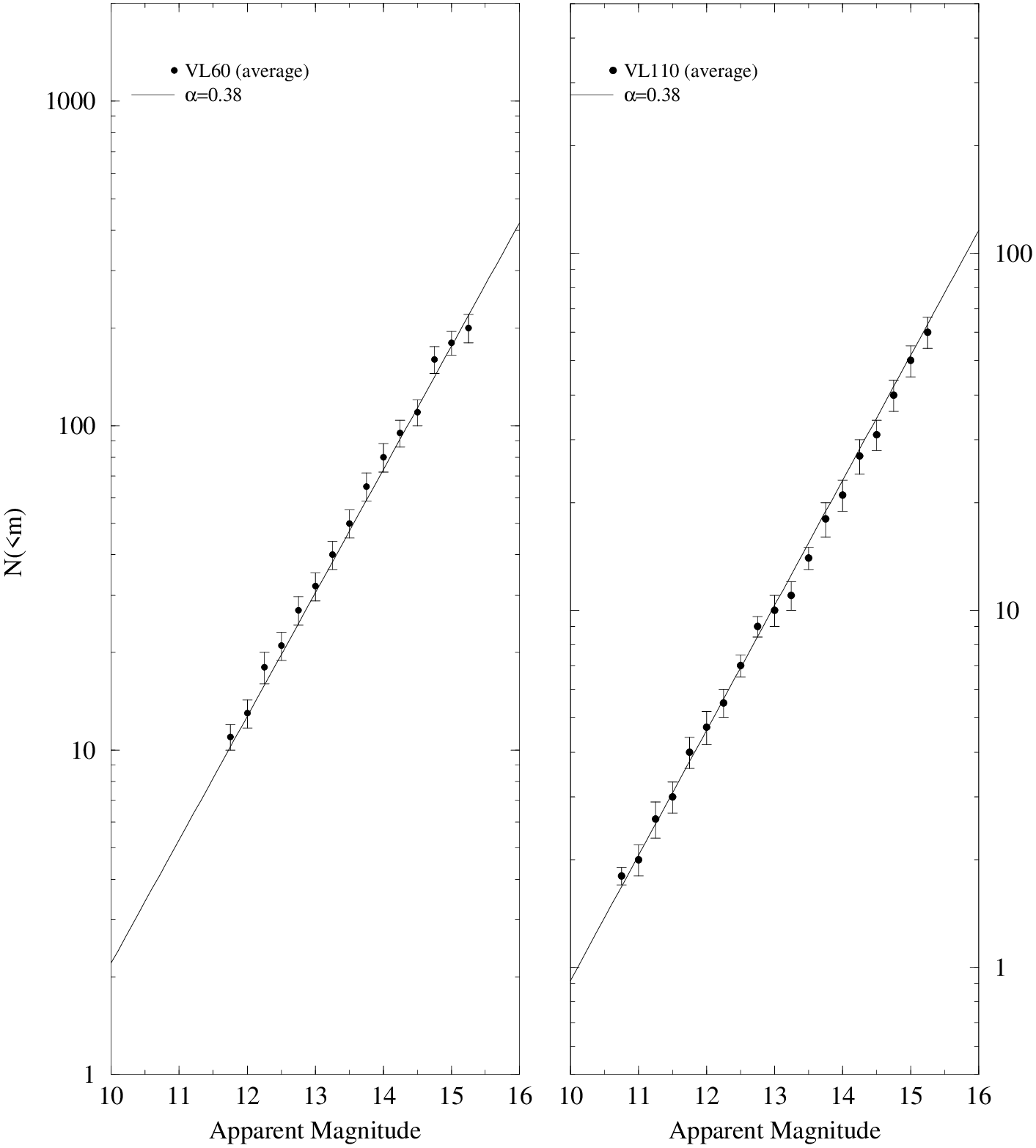}} 
\caption{\label{fig89}  
{\it Left panel}  The average $N(<m)$ in the VL sample VL60. 
 The squares crosses
 refer to  the average $N(<m)$ computed assigning to all the
 galaxies the same absolute magnitude $M_{0}=M^*$. The reference
 line has a slope   $\alpha=0.4$.  {\it Right panel} The average
 $N(<m)$ in the VL sample VL110. The squares crosses refer to  the
 average $N(<m)$ computed assigning to all the galaxies the same
 absolute magnitude $M_{0}=M^*$. The reference line has a slope 
 $\alpha=0.4$ } 
\eef
 the results for VL60 and VL110: a very well
 defined exponent $\alpha=D/5\approx 0.4$ is found in both cases.
 This is in fully agreement with the average space density (the
 conditional average density $\Gamma(r)$) that shows $D \approx 2$
 in these VL samples.

\subsection{Behavior of   galaxy counts in magnitude limited samples}
\label{countsml}

  We are now able to
 clarify the problem of ML catalogs. Suppose to have a certain
 survey characterized by a solid angle $\Omega$  and we ask the
 following question: up to which apparent magnitude limit 
 $m_{lim}$ we have to push  our observations to obtain that the
 majority of the galaxies lie in the statistically significant 
 region ($r \gtapprox  \lambda$) defined by Eq.\ref{v3}.  Beyond
 this value of $m_{lim}$ we should
 recover the genuine properties
 of the sample because, as we have
 enough statistics, the finite
 size effects self-average. From the previous condition for
 each solid angle $\Omega$ we can find an apparent magnitude limit
 $m_{lim}$ 
 so that finally
 we are able to obtain $m_{lim}=m_{lim}(\Omega)$ 
in the following
 way.

In order to clarify the situation, 
we can now compute the expected value of the counts if 
we use the approximation for the behavior of the mass-length relation 
given by Eqs.\ref{equ22}-\ref{equ24}. Suppose, for seek of clarity,
also that 
$\phi(M)=\delta(M-M_o)$, with $M_o=-19$. 
We define
\be\label{eq41}
\lambda= 10^{0.2 (m_{\lambda} - M -25)}
\ee
where $\lambda$ is given by Eq.\ref{v3}.
Then the differential counts are given by
\be
\label{eq42}
\left(\frac{dN}{dm }\right)_i  =  \frac{\log_e 10} {5} 3 B_1 \cdot 
   10^{ \frac{3}{5} m} \cdot 10^{ - \frac{3}{5}  (M_o+25)}
\; \; \; \mbox{if} 
\; \;  m \ltapprox m_{\lambda}
\ee
and 
\be\label{eq43}
\left(\frac{dN}{dm }\right)_i =  \frac{\log_e 10}{5} BD 
  \cdot 10^{\frac{D}{5} m} \cdot 10^{ - \frac{D}{5}  (M_o+25)}
\; \; \; \mbox{if} 
\; \;  m \gtapprox m_{\lambda}
\ee
If $M_o=-19$ and $\lambda \sim \frac{30}{\Omega^{1/D}} (\hmp) $ we have 
\be
\label{eq44}
m_{lim} = m(\Omega) \approx 14 - \frac{5}{D}\log \Omega
\ee

In order to give an estimate of such an effect 
if $\Phi(M)$ has a Schechter like shape, we can 
impose the condition that, in a ML sample, the peak of 
the selection function, which occurs at distance $r_{peak}$,
satisfies the condition
\be
\label{cayy1}
r_{peak} > \lambda
\ee
where $\lambda$ in the minimal statistical 
length defined by Eq.\ref{v3}.
The peak of the survey 
selection function occurs for $M^* \approx -19$ 
 and then we have 
$r_{peak} \approx 10^{\frac{m_{lim}-6}{5}}$.
From the previous relation and from Eq.\ref{cayy1} and Eq.\ref{v3}
we easily recover Eq.\ref{eq44}.

 The magnitude $m(\Omega)$
separates the $0.6$ behavior, strongly dominated by
intrinsic and shot noise fluctuations, from the asymptotic $0.4$
behavior. Of course this is a crude approximation
in view of the fact that $\lambda$ has not a well defined value,
but it depends on the direction of observation and not only on the solid
angle of the survey. However the previous equations gives a 
reasonable description of real data.

We show in Fig.\ref{fig90} the condition given by Eq.\ref{eq44}.
\bef 
\epsfxsize7cm
\centerline{\epsfbox{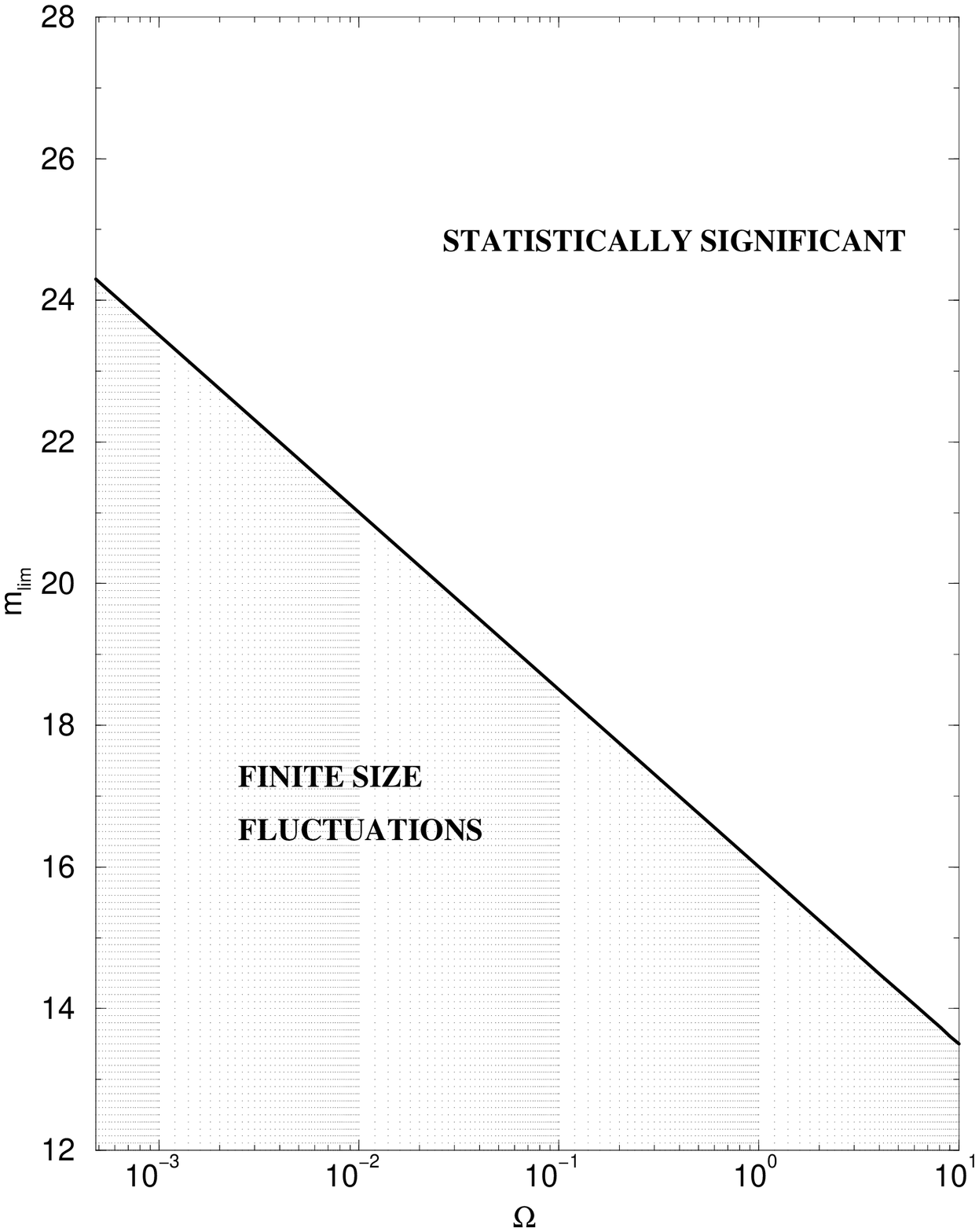}}
 \caption{\label{fig90} If a survey
 defined by  the apparent magnitude limit $m_{lim}$ and the solid
 angle $\Omega$ lie in the  statistically significant  region it is
 possible to obtain the self-averaging properties of the
 distribution also with the integral from the vertex. 
Otherwise one needs a redshift survey which contains the three
 dimensional information, and then one can perform average.
 Only in this way it is possible to smooth out the finite size
 effects. } 
\eef 
From the previous figure it follows that for $m > 19$
 the statistically significant region  is reached for almost {\em
 any} reasonable value of the survey solid angle. This implies that
 in deep surveys, if we have enough statistics, we   readily 
 find the right behavior ($\alpha =D/5$) while it does not happens
 in a self-averaging way for the nearby samples. Hence the exponent
 $\alpha \approx 0.4$ found in the deep surveys ($m>19$) is a {\em
 genuine feature of  galaxy distribution}, and corresponds to real
 correlation properties.  
In the nearby surveys $m < 17$ we do not
 find the scaling region in the ML sample for almost {\em any}
reasonable
 value of the solid angle. Correspondingly the value of the
 exponent is subject to the finite size effects, and to recover the
 real statistical properties of the distribution one has to perform
 an average.  

From  the previous discussion it appears now clear
 why a change of slope is found at $m \sim 19$: this is just a
 reflection of the lower cut-off of the fractal structure and in
 the surveys with $m_{lim} > 19$ the self-averaging properties of
 the distribution cancel out the finite size effects. This result
 depend very weakly on the fractal dimension $D$ and on the
 parameters of the luminosity function $\delta$ and $M^*$ used. Our
 conclusion is therefore that the exponent $\alpha \approx 0.4$ for
 $m > 19$ is a genuine feature of the galaxy distribution and it is
 related to a fractal dimension $D \approx 2$, which is found for $m
 < 19$ in redshift surveys only by 
{\em performing averages}.  We note
 that this result is based on the 
assumption that the Schecther
 luminosity function  holds also at high redshift, or, at
 least   to $m \sim 20$. This result
 is confirmed by the analysis
 of Vettolani \etal \cite{vet94} who found that the luminosity
 function up to $z \sim 0.2$ is in
 excellent agreement with that
 found in local surveys \cite{dac94}.

 Finally a comment on the {\it amplitude} of counts.
In Fig.\ref{fig85} the solid line represents the behavior of
Eq.\ref{q3}. The prefactor $B$ has been determined by the full 
correlation analysis
 while the fractal dimension is $D=2.2$. The
parameter of the luminosity function are $\delta = -1.1$ and $M^*=-19.5$
as usual. The agreement at faint
 magnitudes ($ m \gtapprox 19$) is quite good. At bright
magnitudes one usually 
underestimates the number of galaxies because of
finite size effects, even in some cases the number counts
can be larger than the predicted value. This is 
related to the asymmetry of the fluctuations in a 
fractal structure.
For example we report in Fig.\ref{fig91} various 
determinations of the GNC in different regions of the sky
(from \cite{b96}).
\bef 
\epsfxsize 5cm
\centerline{\epsfbox{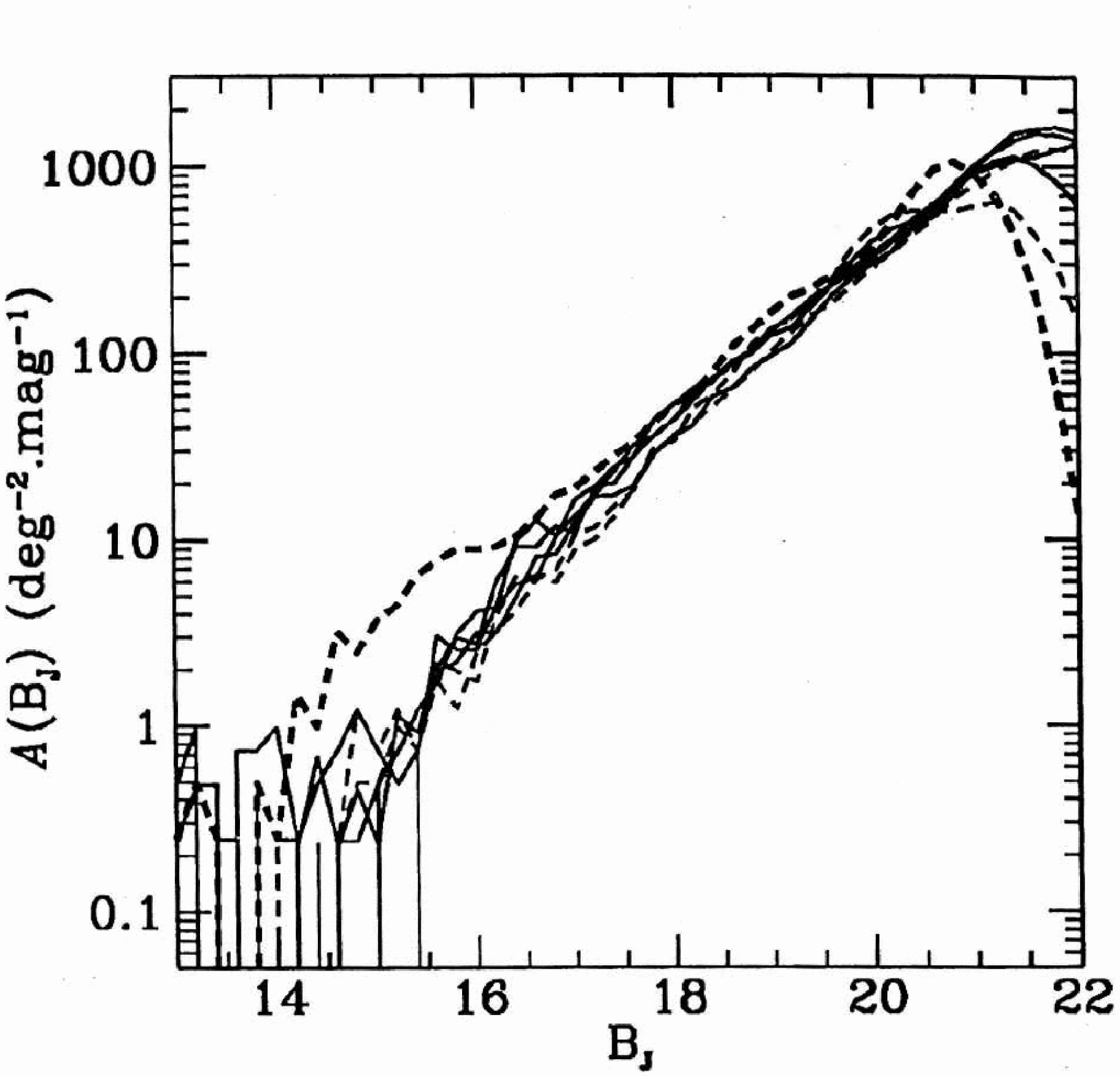}}
 \caption{\label{fig91} Determinations of the GNC in
various sky regions (from Bertin \& Dennefeld, 1996).
 The bright end 
wildly fluctuates around the predicted value, while
at fainter magnitudes the behavior is quite smooth and
is the same for the different fields.} 
\eef
The fact that at faint magnitudes the behavior is quite regular, is due 
to the smoothing of spatial fluctuations for the luminosity effect. 
Namely at a 
given apparent magnitude there are contribution from galaxies located at very
different distance, as the luminosity function of galaxies is spread 
over several decades of luminosities.
 On the other hand at the bright end there 
are contributions only from   nearby galaxies,
 and is such a way the space 
finite size fluctuations are not smoothed out.

In the data shown in Fig.\ref{fig85}  K-corrections have been not 
 applied. 
Such corrections
 must be present because of the Hubble
 distance-redshift relation. Namely
the spectrum of a certain galaxy at 
redshift $z$ is shifted towards
 red  of a certain 
amount, according to 
the Hubble law. There are several galaxies (E/S0) with steep
spectra and hence for
 these one detects a lower value of the intensity of the "true"
one because of the 
shift of the maximum of the spectrum. However there are 
several other galaxies
 (Sdm, Scd) with flat spectra. In some case the 
 shift due to the Hubble law 
may produce an higher intensity 
while in other a decrease of the apparent flux. 
 The K-corrections 
are model dependent corrections: 
the sign can be in both directions,
 i.e. towards an increasing or  a decreasing
of the absolute magnitude \cite{kkcor}. Moreover the 
effect of such corrections is in general  
not so important for the distortion of
 the number counts relation
\cite{yo93}.

 In observational astrophysics there are a lot of data which are
 only angular ones, as the measurements of distances is general a
 very complex task. We briefly discuss here the distribution of
 radiogalaxies, Quasars and the $\gamma$-ray burst distribution.
 Our conclusion is that all these data are compatible with 
a fractal structure with $D \approx 1.6 \div 2.0$ similar to that
 of galaxies.  

\subsection{Radio galaxies}
\label{countsradio}

As an example of counts of objects, we consider the case of radio galaxies.
The majority of  catalog radio sources are
 extragalactic and   some
 of the strongest are at cosmological distances.
 One of the most important information on radio galaxies
 distribution has been obtained from the sources counts as a
 function of the apparent flux \cite{con84}. 
Extensive surveys of sources have been made at 
various frequencies in the range  $  0.4
 \div 5 \; GHz$. In Fig.\ref{fig92} 
\bef 
\epsfxsize 8cm
\centerline{\epsfbox{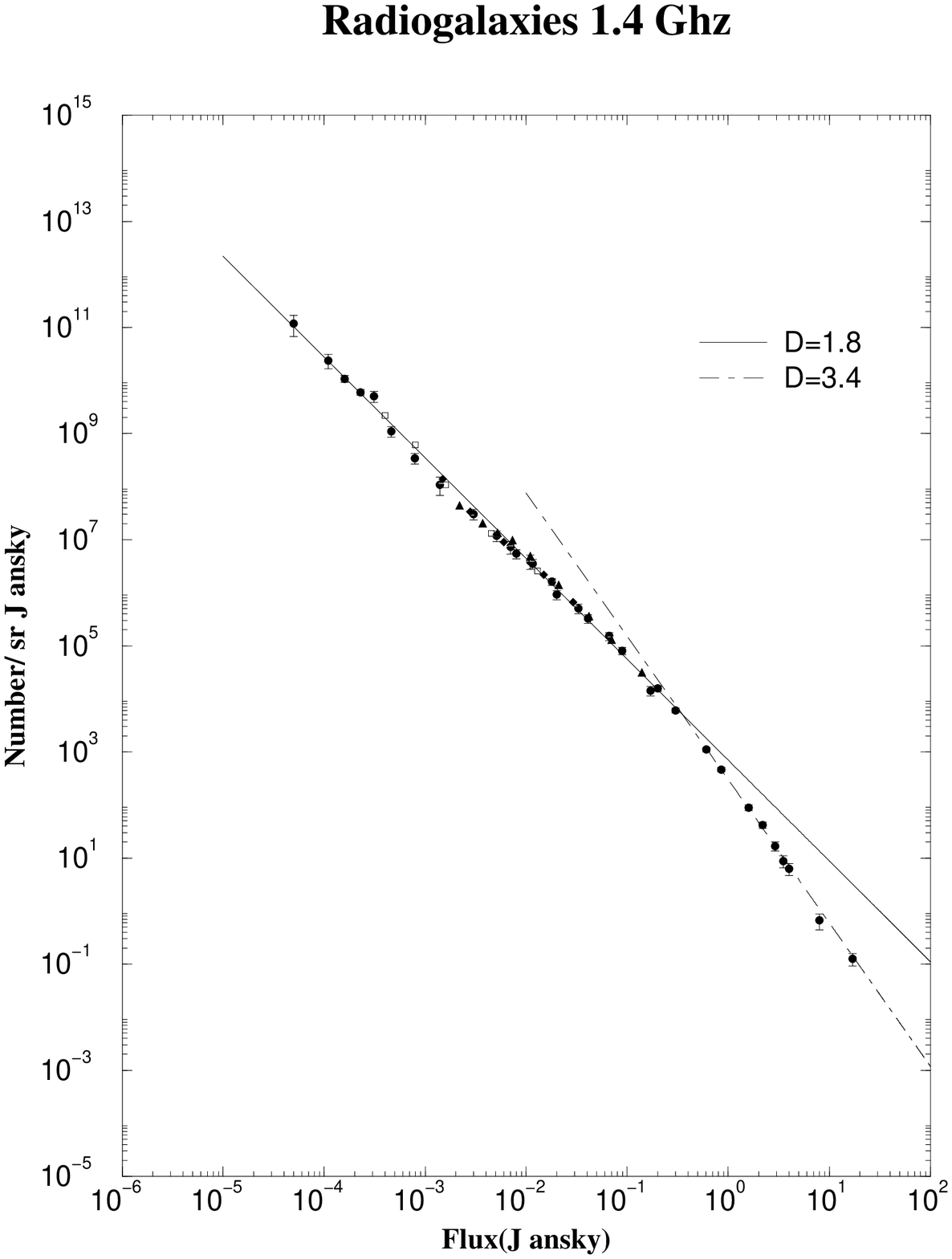}}
 \caption{\label{fig92} Normalized differential sources counts at
 $\nu = 1.4 Ghz$. {\it Abscissa} log flux density (Jy). {\it
 Ordinate}  log differential number of sources $n(S)$. The solid
 line represents the behavior of a fractal  structure with
 $D=1.8$. The amplitude has been computed from the knowledge
of the luminosity function (see text). 
The agreement is excellent, except in the bright fluxes
 region due to the presence 
of finite size fluctuations. (From Condon, 1984). } 
\eef 
we show a compendium of sources counts at $\nu=1.4
 Ghz$ \cite{con84}. The differential counts are plotted against the
 apparent flux. In the bright flux region there is
 a deviation from a power law function, while for four decades the
 agreement with a fractal distribution with  $  D \approx 1.8$ is
 excellent. Such a behavior has been usually explained in
 literature as an effect of sources/space-time
 evolution. Here we propose that
 the radio galaxies are fractally distributed, as galaxies, with
 almost the same fractal dimension. 
The deviation at bright fluxes
 in this picture is explained as a spurious effect due to  the
 small scale fluctuations, as in the
 case of galaxy counts. In the
 other frequency bands the situation is nearly the same
 \cite{con84}. The simple picture of a fractal distribution of
 radio sources is therefore fully compatible with the experimental
 situation. 


It is simple to show that the differential number of 
radio galaxies for unit flux (in Jansky) and 
unit steradian, in an Euclidean space,  is given by:
\be
\label{eqradio1}
n(S)=\frac{1}{(3\cdot 10^9)^2} 
\left(\frac{1}{4\pi}\right)^{\frac{D+2}{2}}
\frac{DB_R}{2} S^{-\frac{D+2}{2}} \int \phi(L) dL
\ee
where we have taken, as usual, 
$n(r) = (B_RD/4\pi) r^{D-3}$ and $S=L/(4 \pi r^2)$
($L$ is the intrinsic luminosity).
From the knowledge of the parameters of the luminosity function 
 \cite{con84} we have that a good approximation of the data
reported in Fig.\ref{fig92} is given by $D=1.8$ and $B_R=0.1 Mpc^{-D}$.
Such a low value for $B_R$ implies that the 
density of radio galaxies is about 100 lower than that of 
optical galaxies. 
When it will be possible to have  a complete redshift 
sample of Radio galaxies, i.e. a well defined volume limited
sample, one will be able to measure directly $B_R$ and $D$ from
the knowledge of the conditional density and hence to compare those
values with the ones we have measured from the  number flux 
relation.

\subsection{Compendium of counts}
\label{countscomp}

In Tab.\ref{tabbande} we summarize the behavior of the number counts
for various astrophysical objects. The small scale exponent corresponds to
the bright end of 
the number counts relation, while the large scale exponent to the 
faint end. The "small scale exponent" {\it is not} a real 
exponent and wildly fluctuates for the different objects. On the 
contrary  the large scale
exponent {\it is} a real
 exponent and its value is in the range $1.8 \ltapprox D 
\ltapprox 2.2$ for almost all the cases. The counts of all these 
different objects is therefore compatible with a fractal distribution in space.

This implies a completely new 
interpretation of the counts. At small scale 
the exponent $D \approx 3$ (i.e $\alpha \approx 0.6$ 
 in the case of magnitude counts)
is due to finite size effects and not to
 a real homogeneity: this has be shown
to be the case with very specific tests.
 On the other hand, at larger scale
 the value $D \approx 2$ 
(i.e. $\alpha \approx 0.4$) corresponds to 
the correct correlation
properties of the samples, that we can find 
by the complete correlation 
analysis in the three dimensional space. 
This implies that galaxy evolution,
 modifications of the 
Euclidean Geometry and the K-corrections 
{\it are not} very relevant 
in the range of 
the present data. In addition the fact 
that the exponent $\alpha \approx 0.4$ 
holds up to magnitude $27 \div 28$ for galaxies seems to indicate that 
the fractal may continue up to distances $\sim 4000 \hmp$. Quite
 a remarkable fact
if one considers 
that the Hubble radius of the Universe is supposed to be  $  4000 \hmp$.
Moreover such a 
behavior can be found for galaxies in the different photometric bands, 
as well as for other astrophysical objects. 

\begin{table} 
\begin{center} 
\begin{tabular}{|c|c|c|c|c|}
 \hline
 &       &            &&   \\ 
Objects     & Small  scale  behavior &  & Large scale behavior &\\
 &       &            &&   \\
 \hline 
 U-band$^{1}$ & $ ? \ltapprox U \ltapprox 18$  & {\bf $D \approx
 ?$}        &  $18 \ltapprox U \ltapprox 24$  &{\bf $D \approx 2.5$}           \\ 
B-band$^{2}$ & $12 \ltapprox B \ltapprox 18$ & {\bf $D \approx 3$}      & $18 \ltapprox B \ltapprox 28$ 
& {\bf $D \approx 2$}              \\ 
V-band$^{3}$ &    $ 
 \ltapprox 20$&  {\bf $D \approx ?$}    & $22 \ltapprox V \ltapprox 25$
 & {\bf $D \approx 1.95$}               \\ 
R-band$^{4}$&     $15 \ltapprox R \ltapprox 18$  &{\bf $D \approx
 2.8$}   &$20 \ltapprox R \ltapprox 26$  &{\bf $D \approx 1.85$}
                \\ 
I-band$^{5}$ &   $? \ltapprox I \ltapprox 19$ & {\bf $D \approx
 ?$}      &  $19 \ltapprox I \ltapprox 25$  & {\bf $D \approx 1.7$}
              \\ 
K-band$^{6}$ &  $12 \ltapprox K \ltapprox 17$ &  {\bf $D \approx
 3.35$}     &  $20 \ltapprox K \ltapprox 24$  & {\bf $D \approx
 1.6$}             \\
&       &            &&   \\ 
Quasars$^{7}$ & $14.75 \ltapprox B \ltapprox 18.75$ &  {\bf $D
 \approx 4.4$}      & $19 \ltapprox B \ltapprox  23$ & {\bf $D
 \approx 1.5$}              \\ 
 Radio$^{8}$ $\nu=1.4 Ghz$ &  $1 \ltapprox S \ltapprox 10$ &{\bf $D
 \approx 3.44 $}      & $10^{-5} \ltapprox S \ltapprox 1$ & {\bf $D
 \approx 1.8$}         \\ 
Radio$^{8}$ $\nu=0.61 Ghz$ &$1 \ltapprox S \ltapprox 10$ &{\bf $D
 \approx 3.8$}        &  $10^{-3} \ltapprox S \ltapprox 1$ & {\bf
 $D \approx 1.6$}              \\
 Radio$^{8}$ $\nu=0.408 Ghz$ &$1
 \ltapprox S \ltapprox 10$ &{\bf $D \approx 3.7$}        & 
 $10^{-3} \ltapprox S \ltapprox 1$ & {\bf $D \approx 1.5$}   
           \\
Radio$^{8}$ $\nu=5.0 Ghz$ &$1 \ltapprox S \ltapprox 10$ &{\bf $D
 \approx 3.4$}        &  $10^{-4} \ltapprox S \ltapprox 1$ & {\bf
 $D \approx 1.8$}              \\
X-ray sources$^{9}$ & $5 \cdot 10^{-13} \ltapprox S \ltapprox
 10^{-12}$ & {\bf $D \approx 3.4$} &  $10^{-16} \ltapprox S
 \ltapprox 5\cdot 10^{-13}$ & {\bf $D \approx 1.8$}              \\
$\gamma-ray$ bursts $^{10}$ & $10 \ltapprox S \ltapprox 100$  &
 {\bf $D \approx 3$}       &  $10^{-1} \ltapprox S \ltapprox 10$  &
 {\bf $D \approx 1.7$}              \\ 
 &       &            &&   \\ 
\hline
\end{tabular}
 \caption{ \label{tabbande} The exponents of counts for different
 kinds of astrophysical objects (see text). (Ref.1-6: see Table 8;
   Ref.7:
 Hartwick  \& Schade,  1990; Ref.8: Condon,1984; Ref.9:  
 Hasinger \etal, 1993; Ref.10: Meegan \etal 1995)  }
 \end{center} \end{table}


\section{Conclusion}

   The highly  irregular galaxy distributions with large structures and 
voids strongly point to a new statistical approach in which the 
existence of a well defined average density is not assumed a priori and 
the possibility of non analytical properties should be addressed 
specifically. The new approach for the study
 of galaxy correlations in all the available catalogues 
shows that their properties are actually compatible with each other 
and they are statistically valid samples. The severe discrepancies 
between different catalogues that have led various authors to consider 
these catalogues as {\it not fair}, were due to the inappropriate methods of 
analysis.
 
The correct two point correlation analysis shows well defined fractal 
correlations up to the present observational limits, from 1 to 
$1000\hmp$ with fractal dimension $D \simeq 2$.
Of course the statistical quality and 
solidity of the results is stronger up to 
$100 \div 200 \hmp$ and 
weaker for larger 
scales due to the limited data. It is remarkable, 
however, that at these larger scales one observes exactly the continuation
of the correlation properties of the small and intermediate scales.

From the theoretical point of view the fact that 
we have a situation characterized by {\it self-similar structures} 
 implies that we should not use concept
 like $\xi(r)$, $r_0$, $\delta N/N$ and certain properties of
the power spectrum, because they are not suitable to represent 
the real properties of the observed structures. 
In this respect also the N-body simulations
should be considered from a new perspective.
One cannot talk about "small" or "large" amplitudes 
for a self-similar structure because of the lack of a reference value like the 
average density.
The Physics should shift from {\it "amplitudes"} towards {\it "exponent" }
and the methods of modern statistical Physics should be adopted.
This requires the development of constructive interactions between two fields.


\section*{Acknowledgemnts}
We would like to warmly thank Prof. N. Sanchez for
useful discussions and for their 
kind hospitality.

\end{document}